\documentclass[12pt]{article}

\begin{document}

\makeatletter
\@addtoreset{equation}{section}
\renewcommand{\theequation}{\thesection.\arabic{equation}}
\makeatother
\def \ci{cite}
\def \const{{\rm const}}

\newcommand{\eq}[1]{(\ref{#1})}

\vspace*{1cm}
Imperial/TP/2-03/31
\vskip0.4truecm

\begin{center}
\vskip 0.2truecm {\Large\bf
 On open superstring partition function       \\
\vskip 0.2truecm
 in inhomogeneous  rolling tachyon background
}
\\
\vskip 0.5truecm
{\bf A. Fotopoulos$^{a,b}$ and A.A. Tseytlin$^{c,a,}
$\footnote{Also at Lebedev Physics Institute, Moscow.}
}
\vskip 0.4truecm
 $^a$Theoretical Physics Group, Blackett Laboratory,
 \\ Imperial College, London SW7 2BZ, U.K.
\vskip 0.4truecm
 $^b$Centre de Physique Theorique,
\\ Ecole Polytechnique, 91128 Palaiseau, France
\vskip 0.4truecm
 $^c$ Smith Laboratory, The Ohio State University,
\\ Columbus OH 43210, USA
\end{center}

\vskip 0.2truecm \noindent\centerline{\bf Abstract}
\vskip .2truecm
We  consider  open superstring partition function $Z$ on the disc
in time-dependent  tachyon background $T= f(x_i) e^{\mu  x_0}$
where the  profile function $f$ depends on spatial coordinates.
We compute $Z$  to  second order in derivatives of $f$
and compare the result  with some previously suggested  effective
actions depending only on  first derivatives of the tachyon
field. We also compute the  target-space  stress-energy
tensor in this background  and demonstrate
 its conservation for the linear profile $f= f_0 + q_i x^i$
  corresponding to a marginal perturbation. We comment on
the  role of the rolling tachyon with the linear spatial
profile  in the decay of an unstable D-brane.

\newpage

\renewcommand{\thefootnote}{\arabic{footnote}}
\setcounter{footnote}{0}




\newcommand{\fig}[1]{Fig.~\ref{#1}}
\newcommand{\lp}{\left(}
\def \del {\partial}
\newcommand{\rp}{\right)}
\newcommand{\blp}{\biggl(}
\newcommand{\brp}{\biggr)}
\newcommand{\ze}{\zeta}
\def \ov {\over}\def \td {\tilde}
\def \Z {{\cal Z}}
\def \ci{\cite}\def \foot{\footnote}
\def \k {\kaPa}
\def \ov {\over}
\def \ha { {\textstyle { 1\ov 2}}}
\def \bi {\bibitem}
\def \la{\label}
\newcommand{\rf}[1]{(\ref{#1})}
\def \m {\mu} \def\n {\nu} \def \s {\sigma} \def \r {\rho} \def \l {\lambda}\def\t {\tau}
\def \a {\alpha} \def\b {\beta} \def\ep {\epsilon} \def \apr {\alpha
'} \def \th {\theta}
\newcommand{\ba}{\bar{\alpha}}
\newcommand{\de}{\delta}
\newcommand{\bpartial}{ \bar{\partial}}
\newcommand{\bz}{\bar{z}}
\newcommand{\bw}{\bar{w}}
\newcommand{\bh}{\bar{h}}
\newcommand{\bL}{\bar{L}}
\newcommand{\bp}{\bar{p}}
\newcommand{\grad}{\bigtriangledown}
\def \be {\begin{equation}}
\def \ee {\end{equation}}
\def \bi{\bibitem}

\def \tdbi  {{\rm TDBI}}
\newcommand{\Pa}{ \int [d \t]}
\newcommand{\e}{\eta}
\newcommand{\Th}{\Theta}

\def \P {{\cal P}} \def \PP {{\rm P}}
\def \four {{\textstyle{1\ov 4}}}
\def\no{{}^\circ \hspace{-0.16cm} {}_{\circ}}
\section{Introduction}

Understanding tachyon condensation and  possible role and
  meaning  of tachyon effective action
in string theory is an old and important problem.
In general, trying to find  an effective
action for tachyon field only  does not seem to make much  sense  since
the scale of masses of an infinite set  of massive string modes is
the same as that of the tachyon mass, and thus
keeping the tachyon while integrating out all other string modes may look
unjustified.
One may hope, however,  that in certain situations
(like in much discussed  examples
of non-BPS D-brane decay  or brane--anti-brane annihilation)
some aspects of  string dynamics can be captured
by an effective  field-theory action involving only
tachyon field (and massless  modes):  all other massive string modes
  may effectively  decouple at a  vicinity of  certain   conformal points.
Reliable  information about  open string  tachyon  effective  actions
should be important,  in particular,   for  current  attempts of cosmological
applications of string theory.

One important message of studies of open string tachyon condensation
is that the  form   of  tachyon effective   action  may  depend
on  a choice of    region in field space where it should be valid.
For example, near the  standard perturbative string vacuum $T=0$ one
may try to    reconstruct   tachyon effective action from string S-matrix
by assuming that the tachyon is the only asymptotic state and by formally
expanding the string scattering amplitudes in powers of momenta
(assuming  some  off shell continuation).
One then gets  (we use  signature $ -+...+$,  $m=0, 1,2 ..., $
     and set $\a'=1$)
\be\la{pe}
L= - \ha   (\del_m  T)^2  +   \ha   \m^2 T^2 -   g_1 T^4 +  ...  \ ,
\ee
where $m^2_{tach}= - \mu^2$.
The linear part of  the  equations of motion  following from \rf{pe} is
the same as the leading-order tachyon beta-function
\be \la{bet}
\del^2_m  T  +  \m^2 T =  0\ , \ \ \ \ \ \ \ \ \
\m^2_{bose}= 1 \ , \ \ \ \ \ \   \m^2_{super} = \ha \ .  \ee
 Since the action \rf{pe}  is reconstructed
from tachyon S-matrix near the tachyon vacuum,
 the  applicability of   local
 derivative expansion  is  doubtful: its form
depends on a particular  ad hoc assumption (not apparently
encoded in the
tachyon  S-matrix) about off-shell continuation.
A separation into derivative-independent and
derivative-dependent terms  is  ambiguous;
 in particular,  the coefficient
of  the derivative-independent  $T^4$ term can
 be changed by a local field redefinition.

One   may  hope to do better by  expanding  near an  end-point \ci{senn,harv}
of tachyon  evolution ($T \gg 1$)  where tachyon  gets ``frozen''
and one may expect that derivative expansion may make more   sense.
Then,   following  \ci{minah,ger,mor},
  one finds  from the derivative expansion of the
superstring partition function  on the disc   \ci{Tse00}
\be \la{deri}
L=   -    e^{-\frac{1}{4} T^2}  \left[1  + \ha    (1 +  b_1 T^2)  (\del_m T)^2  +
   ... \right] \ , \ee
where dots stand for higher-derivative terms
and $ b_1 =  \  \ln 2 - \ha$.
Here  we included $\del^2 T$ contribution  and then integrated it by
parts. The  field redefinition ambiguity
(the  cutoff  dependent  coefficient of $\del^2 T$ term)
was fixed so that  the linear part of the resulting  effective equations agrees \ci{Tse00}
with the linear part of the tachyon beta-function
\rf{bet},  i.e.  like \rf{pe}  this  action reproduces correctly the
value of the tachyon mass near $T=0$.
One may  hope that this action may be used to   interpolate
 between the  regions of  small $T$ (vicinity of
the   tachyon vacuum) and large  $T$
 (vicinity of  a new vacuum where tachyon and other open string modes
 get  frozen).

Extending the action \rf{deri}
to higher orders  in derivative expansion  appears to be  a complicated task.
It is not clear if it  makes sense to sum all terms in derivative expansion
that depend only on the first
 but not on higher derivatives of $T$,
e.g., by evaluating the
string partition function on $T=q_m x^m$ background \ci{mor},
since for finite $q_m$ the linear tachyon background
  is not a solution of the resulting equations of motion.
In general, it is not clear how to interpret a  first-derivative
tachyon  Lagrangian
\be \la{laag}
L= - V(T)\ {\rm K}(\del T) \ ,   \ \ee
 or, in particular,  an
 often-discussed ``tachyon DBI'' Lagrangian  \ci{sen99,lambert,garous}
 (see also  \ci{berg}, cf. \ci{sim})
\be \la{dbi}
    L_{\tdbi} = - V(T) \sqrt{- \det( \eta_{mn} + \del_m T \del_n T ) }
= - V(T) \sqrt{1 + (\del_m T)^2} \ .  \ee
 Indeed, there is no a priori  reason to expect that
higher-derivative terms omitted in \rf{laag} should be small on solutions
of the resulting equations, unless $V(T)\approx $const
(which is not the case at least near $T=0$ where one should get the  tachyon
mass term).\foot{For a discussion of various first-derivative tachyon actions see
 \ci{lamb}.}
This is to be contrasted to the usual relativistic particle action
or DBI action
where higher-derivative ``acceleration'' terms  can be indeed
 consistently ignored
since a  constant velocity  motion  or $F_{mn}=$ const is always a solution.

Alternatively,
one   may  try also  to reconstruct the
tachyon effective action  at  a  vicinity of other  exact  conformal
points, e.g., time-dependent background which should
 represent an exact  boundary conformal theory
\ci{GS,Sen02} (Minkowski version of  Euclidean CFT of  \ci{callan})
\be\la {twoo}
T =  f_0   \ e^{{\mu }  x^0}   +  \td f_0 \ e^{- {\mu }  x^0}  \  .   \ee
Its special case is
the  ``rolling tachyon''  background  \ci{GS,Sen02}
\be\la{mar}
  T=\  f_0   \ e^{{\mu }  x^0}  \  .   \ee
The  disc (super)string  partition function in this background\rf{mar}
was recently  computed in \ci{Lar02},  suggesting
that the corresponding ``potential'' term should  look like
\be\la{lar}
V_0 (T) = { 1 \ov  1 + \ha T^2 } \ .  \ee
A remarkable  observation made in a subsequent paper \ci{Nia03}
is that demanding that a generic  first-derivative  Lagrangian \rf{laag}
should have \rf{twoo} (with $\mu=\frac{1}{\sqrt 2}$ in the superstring case)
as its exact
solution fixes its time-derivative part\foot{Demanding
 only that \rf{mar} is a solution does not
fix the action uniquely (see also  \ci{lamb}, where the suggestion to fix
the form of the first-derivative tachyon action by requiring
that it admits  an exactly marginal static tachyon  background
$T= a \sin \frac{x}{\sqrt 2}$ as its exact solution  was
made). For a discussion of a bosonic string variant of the argument of \ci{Nia03}
see \ci{Smed03}.}
  to be
 \be
\la{niat}
 L = - \frac{1}{1 + \ha T^2 }
 \sqrt{ 1+ \ha T^2 - (\del_0 T)^2  } \ .  \ee
If we assume
that \rf{niat} has a direct Lorentz-covariant
generalization  we are led to
 \be
\la{nia}
 L_{\rm KN}  = - \frac{1}{1 + \ha T^2 }
 \sqrt{ 1+ \ha T^2 + (\del_m T)^2  } =
  - \ha   (\del_m T)^2  + \four  T^2 + ...  \ .  \ee
The action \rf{nia}
does agree with \rf{lar} when evaluated
 on the background \rf{mar}
 or \rf{twoo}\foot{At small $\td T$, the leading quadratic
  terms in this  action
match the first two terms in \rf{pe}, i.e.  like \rf{deri}
it  reproduces the correct value of the tachyon mass near $T=0$.
However, the leading terms in derivative
expansion in  \rf{dbi} are not related to \rf{deri}
by a field redefinition
(which may be attributed to their
different ranges of validity, cf. \ci{Nia03}).
}
and, after a field redefinition
\ci{Nia03},
$ \frac{T}{\sqrt{2}} \to \sinh \frac{\td T}{\sqrt{2}}$,  becomes
\be \la{tdbi}
L =    - \td V ( \td  T) \sqrt{ 1 + (\del_m \td T )^2  } \ , \ \ \ \ \ \ \ \ \
\td V= { 1 \ov \cosh  \frac{ \td T}{\sqrt 2}  }  \ .  \ee
This  seems  to vindicate the TDBI action  \rf{dbi}
 (discussed, e.g., in \ci{walch,lambert}),
 but there are several  questions remaining.


One is the range of validity of the action \rf{nia}.
As was proposed in \ci{Nia03}, this action should be valid for tachyon
fields  which are ``close'' to the exactly marginal background
\rf{mar}, i.e. for $T= f(x^m) e^{\mu x^0}$ where $f$ is  a
slowly changing function. The idea
 of \ci{Nia03} was to
choose a particular direction in space-time, for which
  the exactly  marginal background  is  $T= f_0 e^{\mu x^0}$
  and then to expand in small {\it spatial} momenta  near this
  point. While this prescription may seem not   to be
   Lorentz-covariant,
  one expects that this breaking of Lorentz invariance is
   ``spontaneous'', i.e.  the corresponding effective action
   summarizing dynamics of small perturbations near the  exact conformal
   point can  still be   chosen  Lorentz-invariant.

 Indeed, in  addition to the argument in favor of
\rf{niat} based on having \rf{twoo} as an exact  solution and
having agreement with \rf{lar},  ref. \ci{Nia03} contained also
an apparently independent S-matrix based argument supporting the
existence of an action with first derivatives only that reproduces
the leading (quadratic in
 spatial momenta) terms in the corresponding string
amplitudes computed using an analytic continuation from Euclidean space
expressions.\foot{One expects that the effective action should be
reproducing scattering amplitudes with very special kinematics
where all tachyons have  small spatial momenta,
i.e. moving  very slowly. It is also assumed that one
first expands the string (and field theory)
amplitudes in spatial momenta and then imposes  momentum conservation.}

 The relation of this second argument (assuming  momentum conservation
in all directions in Euclidean space) to the first one referring
to the partition function
\rf{lar} where one does not integrate over the zero mode of $x^0$
 and thus
does not impose momentum conservation in $x^0$ direction is
not obvious  at the moment and may be quite subtle.
  Also, it is not clear a priori
 (independently of  the first argument referring to having \rf{twoo} as
 an exact solution)  why to reproduce the  leading $k_ik_j$ terms
 in the n-point  tachyon scattering amplitudes computed using the
 Euclidean continuation prescription of \rf{nia}
 one needs an action involving all powers of derivatives of $T$:
 while all powers of $\del_m T$ do contribute to the single independent
 coefficient \rf{nia} of the quadratic spatial momentum term
 in each n-point amplitude, that term  may well  be reproduced just by the
 Lagrangian $ L= U_0(T) + U_1(T) (\del_m T)^2 $.
  Here $U_0$ and $U_1$  are power series in $T$ such that the
  corresponding field-theory amplitude matches the leading term
  in the one-shell string theory amplitude
   ${\cal A}_n \sim  c_n k_i k_j + ...$.
 The preference  of \rf{nia} may then be attributed to a specific
 scheme choice allowing  to have \rf{twoo} as an exact
 solution.\foot{
 Assuming, as suggested in \ci{Nia03}, that \rf{nia} applies
 at the vicinity
of the rolling tachyon background \rf{mar}, it may  not be a priori clear
why the first-derivative action like \rf{nia} should be  a useful tool:
all higher-derivative terms ignored in \rf{nia} are of the same order
on this  exponential  background
(and they are small only near $x_0=0$ or small values of $T$
where one can
in any case ignore the non-linear terms in \rf{nia}).
In general, higher-derivative terms (that should again admit \rf{twoo} as
an exact
solution) may be crucial for correctly reproducing string dynamics
(like string fluctuation spectrum) at a  vicinity of
the  rolling tachyon background.
However, as suggested  in \ci{Nia03},
 there should exist a scheme in which
all higher-derivative terms can be effectively traded for
the first-derivative ones.}


Given somewhat indirect nature of the above  arguments it
 would obviously be interesting to support them by explicit
 scattering amplitude computations  and also to establish a precise
 relation  between the Euclidean continuation prescription used in the
 string S-matrix considerations
  and the real-time partition function
 computation \ci{Lar02} leading to \rf{lar}.
 This  was  part of our original motivation in the present paper.
 \foot{Trying to put \rf{nia} as opposed to \rf{niat} on a
 firmer footing is important also  since
  most of the previous discussions
of the tachyon decay were in the  homogeneous (space-independent)
tachyon  case,  and while they indeed  agree \ci{lambert} with
the Lagrangian \rf{niat} depending only on $\del_0 T$, the range
of validity and practical utility  of its inhomogeneous version
\rf{nia} does not seem to be well understood at the moment.}

  In particular, it would be  important  to see if one can
 reproduce the TDBI   action \rf{nia}
{\it directly}  from the {\it string path integral},
just  like one can obtain the BI action $L(F )$ \ci{FT} and derivative ($\del
F$)
 corrections to it  \ci{And88}  by expanding
near  the  conformal point represented by  a  constant
abelian gauge field strength    background.



In general,
the {\it exact} expression for the string partition function
 evaluated
 on a  background  ($x^m=(x^0,x^i)$)
\be
\la{tf}
T(x^0,x^i) =\  f(x^0,x^i) \   \ e^{\mu x^0 }   \   \ee
  should indeed  be Lorentz-covariant, but that need not apply to its
first-derivative  part only.
Note also that the leading-order
condition  of marginality of such background, i.e.
the equation $\del^2_m  T +  \ha   T =0$ \ ($\mu= { 1\ov \sqrt 2}$)  becomes
\be \la{derr} \del^2_i f  - \del^2_0 f  - \sqrt 2 \del_0 f =0  \ ,
\ee i.e. it mixes  first and second derivatives. More precisely,
for  marginal perturbation like $f= f_0 e^{-\nu x^0 + i k\cdot x} $ we
have $ \nu = {1 \ov \sqrt 2 }k^2 + O(k^4)$,
so expanding in $k_i$ or
in spatial derivatives we have $ \del_0 f \sim \del^2_i f \sim
(\del_i f)^2 $.
It is here that the existence of a specific scheme choice
\ci{Nia03} should be important, and  one would like to understand
how this scheme  should be defined in the context of
a standard (real-time)
 perturbative
expansion of  the string  partition function.


One may notice
that the  action \rf{nia}  does not  admit
 \be\la{linn}
 T= ( f_0 + q_i x^i) e^{\frac{x^0}{\sqrt 2}} \ee
 as its exact solution, while this ``nearby''
 background  which solves \rf{derr} is
  expected, as suggested by the finiteness of
   the corresponding partition
  function discussed below,  to be  an exactly marginal perturbation.
\foot{By ``exactly marginal''  here  we mean only that it solves
the beta-function equations to all orders. While  a
non-normalizable nature of this background may be problematic for
a CFT interpretation, such $T$  is a natural counterpart of a
linear vector potential field describing a gauge field  with a
constant strength, i.e. $A= A_m \dot x^m=  \ha F_{mn} x^m \dot
x^n$. Like a constant gauge field it cannot be  regularly expanded
in plane waves.
}

 This does not, however, imply  a contradiction but rather that
 it may  be that
 \rf{linn} is an exact solution in a different scheme  than
 the one implied in  \ci{Nia03} in which \rf{nia} is supposed
 to be valid.\foot{We are grateful
 to D. Kutasov for an explanation that follows.}
 Indeed, we may consider  \rf{linn} as  a linear in $q_i$  approximation
 to an exactly marginal background  which is simply a boost of \rf{mar}
 ($q_m q^m = -\ha$)
$$
 T=  \ e^{q_m x^m } = e^{ \sqrt{\ha  + q_i^2} x^0 + q_i x_i } $$
 \be \la{boo} =\  \big[ 1 + q_i x_i  + \ha q_i q_j x_i x_j
 +  { 1 \ov \sqrt 2} q_i^2 x^0 + O( q^3)  \big]\
 e^{ \frac{x^0}{\sqrt 2}}
\ , \ee
and which  is thus  an exact solution of \rf{nia}.
The two backgrounds \rf{linn} and \rf{boo}
may  then be related by a field redefinition reflecting
change of schemes
in which each of these  backgrounds is an exact solution.
Notice that this  field redefinition should necessarily
involve time derivatives since  it should transform \rf{linn} into
a particular case of \rf{tf}.\foot{An interesting question is
if such a field redefinition is not changing physics:
while \rf{boo} does not describe a D-brane at $x^0\to \infty$,
the background \rf{linn} may be thought of describing a
co-dimension one D-brane, see below.}

\vskip  0.75cm

With the  motivation  to try to understand  better the structure
of the tachyon effective action in the vicinity of
the rolling background
\rf{mar},
and, eventually, a scheme choice in which \rf{nia} should be valid,
we would like
to  compute  the  leading terms  in  derivative  expansion
of the string partition function on the disc
for the
inhomogeneous tachyon
background  \rf{tf} that generalizes \rf{mar}.
We shall mostly consider  the case when
the profile function $f$ in \rf{tf}   depends only on the {\it spatial }
coordinates, i.e.  $ f= f(x_i)$.
Such   background is marginal  (i.e. satisfies \rf{bet}) if
$\del^2_i  f =0$.
As a result,  the expansion of the string partition
function  $Z$ in derivatives  of $f$  seems  as
well-defined as     the  expansion  in derivatives of
 any massless-level scalar mode.

Unfortunately, as we shall see below,  expanding  $Z$
near  $f$=const   it does not
appear  to be straightforward to  sum up all terms depending only
on  the
{\it first} spatial derivative $\del_i T =\del_i f\  e^{\mu x^0} $ of the tachyon.
In general, separation of  the  terms  that depend
only on $\del T$ and not on higher derivatives is {\it ambiguous}
(in particular, in  view of the presence of an
 overall potential factor and a
 possibility to integrate by parts).\foot{
Moreover, one  is used to  think that
 if  $Z$ or the effective action is computed
 in derivative expansion in
$\del_i T \sim \del_i f$,
 one should treat all terms with the same number
of derivatives on an equal footing, e.g.,
$(\del f)^8$ and $(\del^4 f)^2$ should be equally important.
One analogy is  with a   massless scalar (e.g., dilaton)
action in closed string theory: there is no known simple
 way to obtain  a closed action involving only
first derivatives of the dilaton;
moreover, a  summation of all terms with
first derivatives only
would  contradict a
low-energy expansion  which is an expansion in powers of derivatives.
Again,  the  case of the BI action
 is different: there, because of gauge invariance,
the field strength $F$ itself is playing the role
of a fundamental field analogous to a massless scalar  while
 $\del F$  is a  counterpart of a scalar derivative, so $F=$const is always a
 solution.
As a result, summing all orders in $F$
 while ignoring $\del F$ terms makes sense, while
summing all orders in $\del F$ while ignoring all higher $\del^n F$ terms
would  not.}


Our aim  here will be more modest:
starting with the disc string  partition function in
 the background \rf{tf} with $f=f(x_i)$, i.e., in the superstring case,
 \be \la{ggg}
 T= f(x_i) \ e^{{\frac{x^0}{\sqrt 2}}} \  , \ee
we shall  compute  directly   the  {\it first two}
leading terms in  $Z$,  or in the corresponding
effective Lagrangian,   in expansion in number
of spatial derivatives
\be\la{dee}
L = - \Z\ , \ \ \ \ \  \ \ \ \ \
\Z=  V_0 (T) + V_1 (T) (\del_i T)^2 + ...  \ . \ee
Here $\Z$ is the integral density in the partition function $Z$
and   dots stand for terms  with more than two derivatives
(assuming possibility of   integration by parts).
We  shall  use the methods and  results
of  ref.\ci{Lar02}
(confirming   and generalizing some of them), which also
emphasized (along with \ci{Sen02})  the importance
of studying spatially-inhomogeneous tachyon  backgrounds.
We shall complement the  expression \rf{lar} for the ``potential''
function $V_0$  found in
\ci{Lar02} with the one for the ``gradient'' function $V_1$ in \rf{dee}
 \be\la{rezz}
 V_1 =
 \frac{1- \ln(1 + \ha T^2)   }{(1 + \ha T^2)^2 }  \ ,    \ee
where we have set one ambiguous (field-redefinition dependent) coefficient
to zero (see Section 4).

 This looks different than  the  coefficient appearing in \rf{nia}
 upon
 substituting \rf{ggg} into \rf{nia}
 \be \la{knn}
 L_{\rm KN} = - V_0 (T) \big[ 1 + \ha (\del_i T)^2 + ... \big]
 =- \frac{1}{1+\ha T^2}\big[ 1 + \ha (\del_i T)^2 + ... \big]  \ ,   \ee
but as in the case of \rf{linn} vs. \rf{boo}
 that could  be attributed to
a  difference in scheme choices:
the background  that should correspond to \rf{ggg}
(viewed as a background for \rf{nia})  in the standard
perturbative scheme used to compute $Z$ should correspond  to a
particular case of \rf{tf} with {\it time-dependent}
profile $f(x^0,x^i)$. Then  to compare to \rf{nia} we would need to know
also some time derivative dependent terms in \rf{dee} and their
 value   on the corresponding $T$-background.

Given the near-on-shell nature of the background \rf{ggg},
it is natural to interpret  $\Z$ as an effective potential energy
produced on the D-brane by the tachyon profile function $f$.
As we shall find below  by
  explicitly  computing   the
 stress-energy tensor on the background \rf{ggg},
   the energy of the system  also
 changes sign at   finite value of the tachyon field.
The change of sign of the gradient function \rf{rezz}
from positive at $  0 < |T| < T_*$ (where $T_*
 =\sqrt { 2 ( e-1)}  \approx 1.85$) to negative at
$ T_* < |T| < \infty$  which lowers the energy
suggests an instability of the system  appearing at certain moment in time
 (i.e. at large enough  value of $T$ in \rf{ggg})
 -- an instability
towards creation of a spatial inhomogeneity $f \sim x$,\foot{This
may be compared  to the discussion in  \ci{harv,mor} where a relevant
tachyon perturbation $T= q x$ was interpreted as relating
two conformal points $q=0$ and $q=\infty$ through RG evolution, with
 $q\to \infty$
``freezing'' the  $x$-direction and thus
representing a lower-dimensional brane.
Here the evolution happens in {\it real}  time
$ -\infty < x^0 < \infty$, and writing $T= (x-a) e^{\frac{1}{\sqrt2}  x^0}$
 one may interpret the $x^0\to \infty$ region is the one where
 $x$ is fixed at value $a$ (transverse position of co-dimension one D-brane).}
 indicating an  emergence  of  a codimension one  D-brane.
\foot{The reason
for this  sign change in $V_1$  at {\it finite}
 value of $T$ may be related to our neglect of higher-derivative contributions.
The negative contribution of the gradient term to the energy is
suppressed   at
 large times since $V_1 \to 0$ at
$T\to \infty$, so the  energy at the end point
of the evolution  should be finite.
We are grateful to A. Linde  for a discussion of the issue of the
sign change of $V_1$.}


\vskip 0.75cm


The rest of this paper is organized as follows.

We shall start in Section 2 with a  discussion of the bosonic string
partition function in the corresponding  analog  $T= f(x_i) e^{x^0}$
of the background \rf{ggg}. We shall complement the result of \ci{Lar02}
for the homogeneous case $f=$const with the expression
 for the first-derivative
$O((\del_i f)^2)$ term in $Z$.

In Section 3 we shall turn to  the superstring case.
In Section 3.1  we shall
rederive the expression for  $V_0$ in \rf{lar} \ci{Lar02}
for  the ``homogeneous'' (derivative-independent)
  part  of $\Z$  in \rf{dee}. We shall complement the discussion in \ci{Lar02}
by  \ (i) explaining
why  in the particular case of the  background \rf{ggg}
  one can indeed ignore (as was done in  \ci{Lar02})
   the contact $T^2$ term in the  boundary  part of the world-sheet  action,
and
(ii) giving the  general proof of the expression   \rf{lar}  to all orders in $T$
(eq. \rf{lar} was checked in \ci{Lar02} only for the first few orders in expansion
 in powers of $f$).
In Section 3.2 we shall compute the gradient  function  $V_1$ \rf{rezz}
in \rf{dee}.

In Section 4 we  will present the computation
of the stress-energy
tensor in the superstring background \rf{ggg}
to the second order in spatial derivative expansion. We will  show that
 the condition of conservation of the stress-energy
tensor is satisfied  in the case of the
marginal background \rf{linn}.
In Section 5
 we shall  make some further comments on the implications of our
result for $\Z$ \rf{dee} for  the structure of the tachyon
effective action.

Some technical details needed for  the computation of the bosonic
partition function  in the $e^{x^0}$  tachyon background
will be given in Appendix A.  In
Appendix B  we shall discuss a  property of path
ordered integral of a totally  antisymmetric function used in Section 3.
Appendices C, D and E  will provide  some  further  details of the
computation of the integrals appearing in  the superstring case.
Appendix F will contain a list of results used in the technically
involved computation of the stress-energy tensor in Section 4.

\section{Bosonic  string partition function}

Before turning to the superstring case which is our
main goal  it is instructive to compute first the leading terms \rf{dee} in the
partition function in the bosonic string case.

\subsection{General  remarks}

Our starting point will be the open bosonic string path integral
on the disc with the boundary interaction term
\be I_{bndy}=\int {d \t \ov 2 \pi}\ T(x) \ , \ee
where  $T$  is given by \rf{tf}  with $\m=1$  and $f=f(x^i)$.
 For notational
simplicity, we shall sometimes assume that
the spatial  profile function
$f(x^i)$ depends only on   coordinate $x^1\equiv x $.
Expanding the coordinates $x^0, x^1$ near constant (zero-mode)
values  $ x^m \to x^m + X^m (\xi)$ and  writing the
interaction term in the string action as a Taylor expansion in
powers of the fluctuation $X^m$ we get
\begin{eqnarray}\label{Ibnd}
I_{bndy}(x+X)=\int {d \t \ov 2 \pi}
\ e^{x^0+X^0} [f(x) +\  \del f(x) \ X\  +
\frac{1}{2} \del^2f(x)\  X^2\  + ....]\ .
\end{eqnarray}
The general expression for the partition function
is then
\begin{eqnarray}\label{TPF}
Z \sim  \int dx^0 dx\   \Z(x^0,x) \ , \ \ \ \ \
\ \ \ \ \ \ \Z(x^0,x)= <e^{-I_{bndy}(x+X)}>
\end{eqnarray}
Here $ <...>$ is the  expectation value with the free string action on the disc.
 In the superstring case the string partition function $Z$
on the disc  is  directly related to the massless  mode
(gauge vector)
effective action, i.e.  $S[A]$
is equal to  (a renormalized value of) $Z$ computed using $I_{bndy}=
\int d\tau \ A_m (x) \dot x^m$
\ci{tse88,And88};
 the same is expected to be true  also in the tachyon case
\ci{mor,Tse00,prez}.
As for the bosonic case, here the  relation between
$S$ and $Z$ in the tachyon background
 case is less clear a priori;
 an expression for  $S$ (whose derivative should be proportional
 to  the corresponding
  beta-function)
  suggested  within boundary string field theory approach  \ci{wit}
  is\foot{This is also a combination in which linear M\"obius divergence in $Z$
  cancels out \ci{mor} (see also a discussion in  \ci{Tse00}).}
$
S= Z + \beta^T{ \de Z \ov  \de T}\ $,   where
$\beta^T= - T -   \del^2_m T $
is  the tachyon beta function.
 Here  we shall consider only  $Z$, i.e. will not study
   the corresponding  bosonic effective action in detail.

To compute the partition function
  $Z$ by expanding in powers of derivatives of $f$
 will require to know  the correlators with arbitrary
numbers of $e^{X^0}$ insertions and fixed numbers of
$X$-insertions  like
\be\la{piu} \int d\t_1 ...  d\t_n \ < e^{X^0(\t_1)}...
e^{X^0(\t_k)}>< X(\t_{k+1})... X(\t_{n})> \ , \ee
 where respective
correlators are evaluated with respective 2-d free actions, i.e.
 $\int d^2 \xi \
(\del X^0)^2$ and $\int d^2 \xi  \ (\del X)^2$.
 We will  therefore need to use and
extend the methods  of \ci{Lar02} who computed $ \int d\t_1 ...
d\t_n \ < e^{X^0(\t_1)}... e^{X^0(\t_k)}>$.

As a result, we expect to find  (up to a total spatial derivative)
 \be
\la{expa} \Z= V_0 (T) +  U_1(T)  (\del_i T)^2 + U_2 (T)  \del_i^2 T + ...
=  V_0 (T) +  V_1(T)  (\del_i T)^2  +  ...  \ . \ee
Here
$T=T(x^0,x)= e^{x^0} f(x)$ and all derivatives  are over
the
spatial coordinates $x^i$ only, i.e. it is assumed
that $f$ and thus $T$  are
slowly varying in spatial direction.
According to \ci{Lar02},  the homogeneous
($\del_i f=0$) part of the bosonic partition function  is
\be \la{zes}
\Z _{0} = V_0 = { 1 \ov 1 + T} \ . \ee
Our aim  will be to compute $V_1(T)$   in \rf{expa}.


Let us  first  comment on some technical aspects of  the
computation of $\Z$. As usual, one expects to find  2-d
divergences coming from contractions of the fluctuation fields
$X^m$ at the same point; these  can be renormalized by a  standard
field redefinition  of the tachyon field involving the
beta-function  \rf{bet}, i.e.  $ \Z_{ren}=$exp$(\ha  s_1 \beta^T
\frac{\partial}{\partial T}) \Z ,$ where $s_1$ is a cut-off
dependent (i.e.  ambiguous) coefficient. There are no other
divergences from coincident points -- possible divergences coming
from powers of propagators $<XX>$  in \rf{piu}  turn out to be
suppressed by the contributions of the $< e^{X^0}... e^{X^0}>$
correlators. This may look surprising since in the bosonic string
case one expects  also power divergences corresponding to the
M\"obius infinities  in the scattering amplitudes
\cite{tse88,And88}. The M\"obius infinities are effectively
hidden in the remaining
 integrals over the zero modes $x^m$:
not performing integrals over $x^m$
is equivalent to not imposing
 momentum conservation, and this  effectively regularizes the M\"obius
infinities.\foot{
Consider,   for example, a momentum non-conserving two-point function of
the tachyons on the disc, i.e.
$ \int < V_{p_1}(z_1) V_{p_2}(z_2)> \sim \int
\frac{dz}{z^{1-\apr p^2}}$,
where $p_1+p_2=p$. When $\apr p^2 \to 1$ the 2-point function
diverges logarithmically which corresponds to a pole from the
propagator of an intermediate state. For $p^2 \to 0$ the amplitude
diverges linearly with a cut-off  which  represents  the M\"obius
infinity.}


Before turning to the general case,  let  us  consider first
 the   2-point (order $f^2$)
contribution to the two-derivative
term in the partition function
\foot{We have omitted the linear  term $  s_1 e^{x^0} \del^2_i f$
since being a total derivative in spatial directions
it integrates to zero. It will be included in the general
 expression in the next subsection.}
\be\la{C22}
  \Z_2 =   (e^{x_0})^2 [I_2 (\del f)^2  + I'_2 f  \del^2 f ]  + ...  \ , \ee
i.e.
\be \la{zz}
\Z_2=  -(\del_i T)^2  +   s_1   T \del^2_i T + ...
\ .  \ee
We have used \rf{zes} and that
\be\la{tak}
 I_2=\ha \int <e^{X_0(\t_1)}X(\t_1) e^{X_0(\t_2)}X(\t_2)>=
 -\frac{1}{\pi} \int_{- \pi}^{\pi}d\t\  \sin^2\t \ \ln(4\sin^2\t)=
-1\ ,  \ee
\be I'_2= \ha \int <e^{X_0(\t_1)} e^{X_0(\t_2)}X^2(\t_2)>=
- 2 \ln \ep = s_1
\ .  \ee
Here  $s_1= <X^2(\t) >=G(\t,\t)=- 2 \ln \ep$ is a regularized
propagator at coinciding  points.
The coefficient $s_1$ is ambiguous (field redefinition dependent).
Adding $\Z_0$ \rf{zes} and integrating by parts
 gives,  to quadratic order in $T$, \foot{If we define
the effective  action  as  $S= Z + \beta^T {\del Z \ov \del T}$
with the beta-function given  in the present case by
$\beta^T = - \del_i^2 T $, we get, to quadratic order in $T$
and $\del_i T$:\
$
L= 1 - T + T^2   -   2 s_1  (\del_i T)^2 + ... .$
This may  be compared to the standard quadratic terms
in the  bosonic tachyon action  \rf{pe} which for $T$ given by \rf{tf}
with $f=f(x^i)$
becomes  simply  $L= - \ha (\del_i T)^2 $.
 }
\be \la{zzz}
\Z= 1 -  T +  T^2  - (1 + s_1)   (\del_i T)^2    + ...
\ .   \ee
Let us now generalize \rf{zzz}  to   include  all terms
in  expansion in powers of  $T$ but still keeping
contributions  with only two spatial derivatives of $T$.


\subsection{Two-derivative term in  bosonic partition function }

To  compute the complete  two-derivative part of the
partition function  we shall
 use the method of orthogonal polynomials
following   \cite{Lar02} (see also \cite{Gin93}).\footnote{We wish to thank
F. Larsen and A. Naqvi for sharing with us their
unpublished notes on the
computations in \cite{Lar02}. }
Some technical details
of  the computation are given in Appendix \ref{Abos}.

Let us concentrate on  the finite $U_1(T) (\del T)^2$ term in $\Z$
in \rf{expa}.  The $U_2(T) \del^2 T$ term coming from contraction
$<X^2>$  at one point has   divergent (ambiguous) coefficient and
can be eliminated by a field redefinition $ f \to f +
s_1 \del^2 f $ in the potential term  \rf{zes}.
The  finite
$(\del f)^2$ term in $\Z$ is given by: \be \la{Z2B}\Z_{2\ fin} =
\sum_{n=2}^{\infty}\   \frac{(-1)^n }{2 (n-2)!}\  I_n\
 (e^{x^0})^n  \ f^{n-2}\ (\del f)^2
 \ . \ee
$I_n$ is an  integral of the  Vandermonde determinant \ci{Lar02}
 \be \la{van}
\Delta(\t) = \prod^n_{i<j=1} (e^{i\t_i}-e^{i\t_j}) \ ,  \ee
coming from $  < e^{X^0(\t_1) }...e^{X^0(\t_n)}>$
with an insertion of   logarithmic  $<X X>$ propagator:
\be\la{Mn2}  I_n =- \int \prod_{i=1}^n \frac{d\t_i}{2 \pi}\
|\Delta(\t)|^2\   \ \ln[4\sin^2(\frac{\t_1-\t_2}{2})] \ . \ee If
we expand the logarithm in \rf{Mn2}   in terms of cosines (as, e.g.,  in
\ci{FT})
 we can show that:
\be\la{show}
 I_n = 2\int
\prod_{i=1}^n \frac{d\t_i}{2 \pi} \  |\Delta(\t)|^2 \
 \sum^\infty_{m=1} \frac{1}{m} \cos m(\t_1-\t_2)= -2 (n-2)! \
\sum_{m=1}^{n-1} \frac{n-m }{m}\ .  \ee The final result  for the
all-order  form of the $(\del f)^2$ term in \rf{C22} is given by
substituting $I_n$ into  (\ref{Z2B}): \be \Z_{2\ fin} =
-\sum_{n=2}^{\infty}\ (-1)^n (e^{x^0})^n \ f^{n-2}\ (\del f)^2
\sum_{m=1}^{n-1} \frac{n-m}{m} \ .  \ee The $n=2$ term here agrees
with \rf{C22}. Observing  that the resulting double sum can be
written as a product of two series we finally get  for the  finite
part of $\Z_2$ (with $T= e^{x^0} f(x^i)$):
 \be \la{finn} \Z_{2\ fin} = -\frac{\ln(1+T)}{T
(1+T)^2 }(\del_i T)^2\ .  \ee Including also  the
derivative-independent term $V_0$ \rf{zes}  and the divergent
$\del^2 f$ term (given by $ \ha s_1 \frac{\partial V_0}{\partial T}
\del^2_i T$)
  we finish with
\be \la{zer}
\Z = { 1 \ov 1+ T} \left[
1  - \frac{\ln(1+T)}{T (1+T) }(\del_i T)^2 -  { s_1 \ov 2( 1 + T)} \del^2_i T
+ ... \right]
\ ,    \ee
or, after integration by parts,
\be \la{zerk}
\Z = { 1 \ov 1+ T} \left(
1  - \bigg[ \frac{  \ln(1+T)}{ T (1 + T)}
  +     \frac{s_1  }{ (1+T)^2  } \bigg]
 (\del_i T)^2
+ ... \right) \ .    \ee Expanded at small $T$ this reduces to
\rf{zzz}.




\section{Superstring  partition function}

In the open superstring case  the effective action
should be directly  equal to the  (renormalized) disc partition function.
Our aim here will be to compute the two leading terms \rf{dee} in
spatial derivative expansion of  the action
  by evaluating the superstring partition function
$Z$
in the background \rf{tf} (with $\m=\m_{super}= { 1 \ov \sqrt 2}$)
or \rf{ggg}.

The derivative-independent  part of $Z$ was found  in \ci{Lar02}
to be equal to \rf{lar}.  There were some minor  gaps in the derivation
(cf.  eqs.(61) and (67) and footnotes 6,7  in \ci{Lar02})   which we shall
fill in below in Section 3.1 (some technical details will be
explained also in Appendices B,C,D).
In section 3.2 we shall compute the second-derivative term in $Z$,
obtaining the analog of the bosonic expression in \rf{zerk}
(see also Appendix E).

\subsection{Derivative-independent (``potential'') term }

The starting point is the world-sheet  supersymmetric expression for
the tachyon coupling in the open NS  string
\ci{wik,harv,mor}
(see also \ci{Tse00,krl})
\be
\la{sut}
I_{bndy}
= \int {d\t\ov 2 \pi}  d\th\  [\hat{\ze} D \hat{\ze} + \hat{\ze} T(\hat x) ]=
 \int  {d\t\ov 2 \pi} \ [ \ze \dot{\ze} + \ze \psi^m \partial_m T (x)
+ h T (x)  + h^2 ] \ ,
\ee
where $\hat \ze $ and $\hat  x^m$ are 1-d superfields
with components $\ze,h$ and $x^m, \psi^m$.
Integrating out $h$ we are left with
 \be
\la{put}
I_{bndy} = - { 1 \ov 4}\int   {d\t\ov 2 \pi} ( T^2 - \  \psi^m \partial_m T \  \partial^{-1}_\tau \ \psi^n \partial_n T ) \ .
 \ee
Computing  the path integral in the supersymmetric form starting with \rf{sut}
as in, e.g.,  \ci{And88} we will  need to use
$$
<\hat{\ze}(\t_i,\th_i)  \hat{\ze}(\t_j,\th_j)>= \hat \Theta(i,j)\equiv
\Th(\t_i - \t_j + \th_i \th_j) = \Th(i,j) +
   \th_i \th_j\  \de( i,j) \ ,
$$
\be \la{opi}
\Th(i,j)\equiv \Th(\t_i - \t_j) \ , \ \ \ \ \ \ \ \ \ \ \ \
\de(i,j)\equiv \de(\t_i - \t_j) \ ,
  \ee
where $\Th(\tau)$ is a step function.

Starting with
\rf{sut}  with the tachyon coupling  $T =  e^{x^0\ov \sqrt{2}}\ f(x) $ we find for
 the  $\del f$  derivative-independent term in the superstring  partition function
$$\Z_0= \sum_{n=0}^{\infty} (-1)^n (e^{x^0\ov \sqrt{2}})^{2n} f^{2n}
\int \prod_{i=1}^{2n} \frac{d \t_i}{2 \pi} d \th_i \
\hat \Th(1,2)\dots
\hat \Th(2n-1,2n)  $$
\be \la{zse} \times
 \prod^{2n}_{i<j} |e^{i\t_i}-e^{i\t_j} -i e^{i(\t_i+\t_j)} \th_i \th_j|
\ .  \ee
One can show that here
the contact $\de( i,j )$ part of the
supersymmetric theta-functions
 drops out  since it
gives zero whenever it is picked up by the $d \th_i$ integrations.
Equivalently,
 the ``contact'' $T^2$-term in the component form
of the boundary interaction term  \rf{put}
can be omitted  since it does not contribute to the final result.
 This explains  why the result of
\cite{Lar02}  where the $T^2$-term was not included from the very
 beginning  is indeed   the same as obtained using the
manifestly    world-sheet supersymmetric
boundary interaction \rf{sut}.

Integrating over  $\theta_i$ we finish with the following
expression for the coefficient
of the $n$-th  term in the sum \rf{zse}:
\be\la{SPF}  J_n= \int  [d\tau]_{2n}
\sum_{\P} (-1)^{\P(i_1,\dots,i_{2n})} \ep(i_1,i_2)\dots\ep(i_{2n-1},i_{2n})
\frac{\prod^{2n}_{i<j} G(i,j)}{G(i_1,i_2) \dots G(i_{2n-1},i_{2n}) }\ee
\be \la{mea}
    [d\tau]_{2n} \equiv \prod_{i=1}^{2n} \frac{d \t_i}{2\pi}\
 \Th(1,2) \dots   \Th(2n-1,2n) \ . \ee
We used that $ |e^{i\t_i}-e^{i\t_j} -i e^{i(\t_i+\t_j)} \th_i
\th_j| = |e^{i\t_i}-e^{i\t_j} |  + \ep( \t_i -\t_j) \th_i \th_j$,
where $\ep(\tau)$ is the  sign function.

Equivalently, \be \la{eqq} J_n=\int  [d\tau]_{2n}
\prod^{2n}_{i<j}\ep(i,j)\  W(1,..., 2n) \ , \ee
 \be \la{www}
W(1,..., 2n) \equiv \sum_{\P} (-1)^{\P(i_1,\dots,i_{2n})}
\frac{\prod^{2n}_{i<j} D(i,j)}{D(i_1,i_2) \dots D(i_{2n-1},i_{2n})} \ .
\ee Here we used the following notation: \be\la{not}
\ep(i,j)=\ep(\tau_i-\tau_j)=\Th(i,j)-\Th(j,i) \ , \  \ \ \ \ \ \
G(i,j)=|e^{i\t_i}-e^{i\t_j}|= \ep(i,j) D(i,j) \ , \ee \be \la{der}
 D(i,j)=i[ e^{i(t_i-t_j)/2}-e^{-i(t_i-t_j)/2}] = 2 \sin(\frac{t_i-t_j}{2}) \ . \ee
$\P(i_1,\dots,i_{2n})$  means all $(2n-1)!!$ permutations of
ordered pairs  of the $2n$ indices.\foot{For example,
 in the case
of $n=3$  we have schematically (with sign factors included):
 $\{ (1,2);(3,4);(5,6)\} \to (-)\{
(1,3);(2,4);(5,6)\}\to (+)\{ (1,4);(2,3);(5,6)\}$, etc.}

One  can check that
$W(1,2,\dots,2n)$  in   \rf{eqq} is symmetric under all interchanges of
the  arguments.
The factor $\prod_{i<j}\ep(i,j)$   is of course antisymmetric.
In Appendix \ref{AP} we show that for a
totally antisymmetric  function $A(1,...,2n)\equiv A(\t_1,... , \t_{2n})$
there is the following relation
\be\la{iden}
\int  [d\tau]_{2n}\
  A(1,\dots,2n)\ = \frac{1}{(2n)!} \int \prod_{i=1}^{2n} \frac{d \t_i}{2\pi}
\prod_{i<j} \ep(i,j) \ A(1, \dots,2n) \ . \ee
Then combining (\ref{SPF})
and (\ref{iden}) we can get rid of the $\ep(i,j)$ factors  and find
\be\la{SPF1}
 \Z_0= \sum_{n=0}^{\infty} \frac{(-1)^n}{(2n)!}
(e^{x^0\ov \sqrt{2}})^{2n}f^{2n}\  \int\prod_{i=1}^{2n} \frac{d
\t_i}{2 \pi}\  W(1,2, \dots, 2n)\ . \ee Finally, using  eq.
(\ref{Z02})  of Appendix \ref{AZ}, i.e. $\int \prod_{i=1}^{2n}
\frac{d \t_i}{2\pi}\ W(1,\dots,2n)= (2n-1)!! \ n!$,
 we can show
that the total  result for $\Z_0$  is indeed the one \rf{lar}  of   ref.\cite{Lar02}, i.e.
\be\la{zoo}
\Z_0 = V_0(T)= { 1 \ov 1 + \ha T^2 } \ , \ \ \ \ \ \
\ \ \ \ \ \   T= e^{x^0\ov \sqrt{2}}   f(x)  \ .
\ee

\subsection{Two-derivative (``gradient'')  term  }

Next, let us compute the two-derivative term in $\Z$
and thus  in the effective action.
Expanding the boundary interaction term \rf{sut}
near  constant  values of the coordinates,
 $\hat x^m = x^m + \hat X^m$, where  $\hat X^m(\t,\theta) $
is a fluctuation superfield,   one has
\be \la{mm}
   \int {d\t \ov 2 \pi}  d\th\  \hat{\ze}\  T(\hat x) =\int {d\t \ov 2 \pi}  d\th\  \hat{\ze}\
e^{x^0+\hat X^0  \ov \sqrt{2}}\
\big[f(x)+ \hat X  \del f(x)+
\frac{1}{2} \hat X^2 \del^2f +...\big] \ . \ee
As in the bosonic case,  the contraction  of the
two $\hat X$-fields  at the same point  produces a  logarithmic divergence
that  can be renormalized  away by a redefinition of the
tachyon coupling, i.e. the $\del^2f$ term enters $\Z$  with an ambiguous coefficient:
\be\la{aqw}
\Z_{2\ amb.}  =\ha  s_1  \frac{\partial\Z_0}{\partial T} \del^2_i T  \ . \ee
After using \rf{zoo}   and integrating by parts we get  (cf. \rf{zer},\rf{zerk})
\be \la{we}
 \Z_{2\ amb.}  =  \ha
s_1 \frac{1- \frac{3}{2}  T^2}{(1 + \ha T^2)^3}     (\del_i T)^2  \ . \ee
The finite (unambiguous) part of $\Z_2$ is given by
 \be\la{Z1A}
\Z_{2\ fin} = -\sum_{n=1}^{\infty} \frac{(-1)^n}{(2n)!} \ J_n \
(e^{x^0\ov \sqrt{2}})^{2n}f^{2n-2}(\del_i f)^2  \ , \ee
$$
J_n =
 \int \prod_{i=1}^{2n} \frac{d \t_i}{2 \pi} d \th_i \
\hat \Th(1,2)\dots
\hat \Th(2n-1,2n)   $$
\be\la{cow}
 \times \prod^{2n}_{i<j} |D(i,j)|\  \sum^{2n}_{k<l} \blp \ \ln|D(k,l)| +
\frac{\ep(k,l)}{|D(k,l)|}\th_k \th_l \brp \ .
 \ee
There are two types  of terms in the integrand of the
above expression. The first
one involving logarithms can be shown to be antisymmetric,  and then
using the relation in Appendix \ref{AP} we can
replace the path ordered integral by an ordinary integral.
The second one
requires a little more work but after doing the  $d\th_i$ integrations it
turns out to be  proportional to the integrand of $\Z_0$ in \rf{SPF},
 and   at  the end  one gets again an ordinary  integral.
After some manipulations and using  symmetry of the integrand
we find
\be\la{Z1B} \Z_{2 \ fin}  = -\sum_{n=1}^{\infty} \frac{(-1)^n}{(2n-1)!} \ C_n \
(e^{x^0\ov \sqrt{2}})^{2n}f^{2n-2}(\del_i f)^2 \ ,  \ee
\be\la{coe}
C_n= \frac{1}{2n} J_n=  \int \prod_{i=1}^{2n} \frac{d \t_i}{2 \pi}\
      W(1,2,\dots,2n)\ \bigg[ 1 + (2n-1) \ln|D(1,2)| \bigg] \ .
\ee
Expanding
 the logarithm in \rf{coe} in a power series of  cosines
 (as in the bosonic case in \rf{show})  gives
\be \la{Z1C}
C_n=     \int \prod_{i=1}^{2n} \frac{d \t_i}{2 \pi}\
      W(1,2,\dots,2n)\    \left[1  -(2n-1) \sum_{m=1}^{\infty}
\frac{1}{m} \cos m(\t_1-\t_2)  \right] \ . \ee
 The first term  in
the square brackets  gives the same integral as in the previous
subsection (cf. \rf{SPF1}). As explained in Appendix \rf{EE}, the
second  term is also  evaluated following the same reasoning as  in the
zero-derivative case: one is to  expand the cosines in polynomials of
exponents and use the relations of  Appendix \ref{AZ}. We find
\be \la{In} C_n = (2n-1)!!\  (n-1)!  \ (  \sum_{m=1}^{n-1}
\frac{n-m}{m}\ +n ) \ .  \ee One can check that the direct
calculation of \rf{Z1C} for $n=1$ and $n=2$ gives the values
$C_1=1$ and $C_2=9$ respectively,  in agreement with \rf{In}.
Finally, after similar manipulations as in the bosonic case we get
for the finite
 part of $\Z_2$:
\be \la{Zkk}  \Z_{2\ fin} = \frac{1}{(1 + \ha T^2)^2 } \bigg[ 1-
\ln(1 + \ha T^2) \bigg]   (\del_i  T)^2    \ . \ee
For example, the
first two terms of the  small $T$ expansion of \rf{Zkk}
are:
$(\del_i  T)^2 - \frac{3}{2} T^2 (\del_i  T)^2$,
 in agreement with \rf{Z1B} with  $C_1=1$ and $C_2=9$.

Adding  the
potential \rf{zoo} and the ambiguous \rf{we} terms
to \rf{Zkk}  we end up with
the  central result of this paper -- the
 expression for the superstring
partition function to the second
order  in spatial derivatives of the tachyon:
\be \la{SF} \Z=
\frac{1}{1 + \ha T^2 } \blp 1 + \frac{1}{1 + \ha T^2 }\bigg[ 1-
\ln(1 + \ha T^2)   + \ha  s_1 \frac{1 - \frac{3}{2}T^2 }{ 1+ \ha T^2 }
\bigg] \ (\del_i T)^2 + \dots \brp \ .   \ee
This may be compared  to  the bosonic string result \rf{zerk}.

\subsection{Including $x^0$ dependence in  $f$  }
As was already mentioned in the Introduction,
in the case of $f=f(x^i,x^0)$ which is close to a
marginal perturbation (which should satisfy \rf{derr})
 the expansion  in derivatives
should be  organized so that to take into account that
$ \del_0 f \sim \del_i^2 f
 \sim (\del_i f)^2$.
Indeed, expanding  in time derivatives of $f$
we  get
\be \la{Zkko}  \Z_{ fin} = V_0(T)
 +  V_1 (T)   (e^{\frac{x^0}{\sqrt 2}})^2 (\del_i  f)^2   +  K_1(T) e^{\frac{x^0}{\sqrt 2}} \del_0 f
 +  O((\del_i f)^4,
(\del^2_i f)^2,  (\del_0 f)^2, \del_0 f (\del_i f)^2 )
  \ . \ee
Note that expansion in time derivatives of $T$ does not make
sense: using that
 $ e^{x^0 \ov \sqrt 2}
\del_0 f = \del _0 T - { 1 \ov \sqrt 2} T$ one concludes that  if
one expands in $\del_0 T$, then coefficients of  lower-derivative
terms receive contributions from all higher-derivative $\del^n_0
f$, etc.,  terms.\foot{In fact, expansion in  powers of time derivatives of $f$
is also not well defined: because of time-dependent
 coefficients in \rf{Zkko}, terms with  different powers of $\del_0$ may mix in the
 resulting equations of motion.}


Computing $K_1$ in \rf{Zkko} one gets:
 \be \la{hoh} K_1 =- \sqrt{2}
T V_1 (T) \ . \ee Similarly, one can compute with some effort the
coefficients of the next-order terms $(\del_i f)^4,$ $ (\del^2_i
f)^2,$ $  (\del_0 f)^2,$ $ \del_0 f (\del_i f)^2 $. They happen to
contain second powers of $\ln (1 + \ha T^2)$, which may be
prompting a possibility of some resummation of the derivative
expansion.

\section{Stress-energy tensor}

In this section we will evaluate the target-space
 stress-energy tensor (SET) in the  superstring background
\rf{ggg} to  first order in spatial derivative of $f$.
For this
 we will need to compute the expectation value of the
graviton vertex operator in the background \rf{ggg}.

As was  explained  in \ci{Lar02},  the SET can be found from the
following expression \be \la{SE1} T^{mn }= K [  \Z(x^k) \eta^{mn }
+ A^{mn }(x^k)] \ ,  \ee where $K$ is an overall normalization
constant, $\Z$ is the partition function density and \be\la{aaa}
A^{mn} \equiv <:V^{mn}(0,0): \ e^{ -I_{bndy}}>= W^{mn} + \eta^{mn}
\Z(x^k)\ .  \ee Here the graviton vertex operator is fixed on the
center of the  disc\foot{Fixing the position of the graviton
vertex can be always done in the conformal background using
M\"obius symmetry. Away from conformal points this
 represents a particular ``off-shell'' definition of the stress
 tensor.}
  and has
the form:
\be \la{V} V^{mn}= 2 \int d \th d\bar{\th} D \hat X^m(0)
\bar{D}\hat X^n(0)\ .  \ee
We shall define as in
\ci{Lar02} the following modified normal ordering:
 \be \la{NO} \no V^{mn} \no =
 :V^{mn}:-  \eta^{mn} \ . \ee
Then
\be \la{defW} W^{mn}\equiv  {1\ov 2}< \no V^{mn}(0,0) \no\
 e^{ -I_{bndy}}> \ ,  \ee
and so  \rf{SE1} can be written as
 \be \la{SE2}
T^{mn}= 2K\big[ \eta^{mn} \Z(x^k) + W^{mn }(x^k)\big] \ . \ee
We will begin with the computation of the $W^{0i}$ component which
is the easiest:
$$ W^{0i}=
\sum_{n=0}^{\infty} (-1)^n
(e^{x^0 \ov \sqrt{2}})^{2n} \int d\m_n <   \no  \int d \th d
\bar{\th} D \hat X^0(0) \bar{D}\hat X^i(0) \no   $$
\be  \la{W0i1} \times  \prod_{l=1}^{2n}
e^{\hat X^0(z_l) \ov \sqrt{2}}\big[ f^{2n} + f^{2n-1} \del_i f
\sum_{m=1}^{2n} \hat{X}^i(z_m) + \dots\big]>\ , \ee
 where
\be d\m_n=\int \prod_{i=1}^{2n} \frac{d \t_i}{2 \pi} d \th_i \
\hat \Th(1,2)\dots \hat \Th(2n-1,2n) \ee
We  use the results of \ci{Lar02} for the SET
 computation  in the $f$=const background.\footnote{We have also
verified equations (67) in \ci{Lar02} for all $n$ using our method.}.
Only the second term in the parentheses contributes and we
 are lead
to the result
\be \la{W0i2} W^{0i}= \sum_{n=0}^{\infty} (-1)^{n+1}
(e^{x^0 \ov \sqrt{2}})^{2n} \frac{\sqrt{2}}{2^n} f^{2n-1} \del_if
= \frac{1}{\sqrt{2}} \frac{T \del_i T}{1+ \frac{T^2}{2}}\ . \ee
Inserting this into
 \rf{SE2} we get
  \be \la{T0i} T^{0i}=
2K  W^{0i} = \sqrt 2K  \frac{T \del_i T}{1+ \frac{T^2}{2}} \ . \ee
Next, let us compute $W^{ij}$:
$$ W^{ij}=
\sum_{n=0}^{\infty} (-1)^n (e^{x^0 \ov \sqrt{2}})^{2n} \int d\m_n
<\no  \int d \th d \bar{\th} D \hat X^i(0) \bar{D} \hat
 X^j(0) \no
 $$ \be \la{Wii1}
 \times \prod_{l=1}^{2n} e^{\hat X^0 (z_l) \ov \sqrt{2}}\big[ f^{2n} +
f^{2n-1} \del_af \sum_{j=1}^{2n} \hat{X}^a(z_j) + f^{2n-2} \del_a
f \del_b f \sum_{l<m } \hat{X}^a(z_l)\hat{X}^b(z_m) \ee  $$ +\
{1\ov 2} f^{2n-1} \del_a \del_b f \sum_{m=1}^{2n} \hat{X}^a(z_m)
\hat{X}^b(z_m) + \dots \big]>  \ . $$ The first two terms in the
square brackets
 do not contribute.
  The remaining integrals are again of the same type
as in \ci{Lar02} and lead to the expression \be \label{Wii2}
W^{ij}= \sum_{n=1}^{\infty} (-1)^n (e^{x^0 \ov
\sqrt{2}})^{2n}{f^{2n-2} \ov 2^n} \big[ \eta^{ij}(2-2n) (\del_k
f)(\del_k f) + 2n f \del_i \del_j f \big]\ . \ee
Plugging this
into   \rf{SE2} and using the partition function expression
\rf{SF}  from the previous section  we get
$$ T^{ij}= 2K  \bigg[\eta^{ij}{1 \ov 1+ \ha T^2 } + \eta^{ij}(\del_k
T)^2 \blp { 1
 - \ln(1+\ha T^2 )
\ov (1+\ha T^2 )^2} + \ha {s_1} { 1-{3\ov 2 } T^2  \ov (1+\ha T^2
)^3} \brp$$

\be \la {Tii2} + (\del_i T)(\del_j T)  {- \ha T^2 \ov (1+\ha T^2
)^2}+ T(\del_i \del_j T){- 1 \ov (1+\ha T^2 )^2}+ \dots \bigg] \ .
\ee where $T= e^{x^0 \ov \sqrt{2}} f(x_i)$.

It is straightforward to check that the above two components
$T^{0i}$ \rf{T0i} and $T^{ij}$ \rf{Tii2} satisfy
the SET  conservation law
\be \la{con}
\del_0 T^{0i} + \del_j T^{ji}=0
+ O((\del T)^3) \ .
\ee
Here only the first term in  $T^{ij}$ in \rf{Tii2}
is taken into account
($(\del_iT)^2$ terms lead to $O((\del T)^3)$
corrections which we ignore).

The computation of $W^{00}$ turns out to be
 long and complicated so we
will only outline some basic steps   and leave details for
Appendix F. We start with
$$   W^{00}=
\sum_{n=0}^{\infty} (-1)^n (e^{x^0 \ov \sqrt{2}})^{2n} \int d\m_n
<
  \no  \int d \th d \bar{\th} D \hat X^0(0) \bar{D}\hat X^0(0) \no
 $$ \be \la{W001}
\times  \prod_{l=1}^{2n} e^{\hat X^0 (z_l) \ov \sqrt{2}}
\big[f^{2n} + f^{2n-1} \del_i f \sum_{m=1}^{2n} \hat{X}^i(z_m) +
f^{2n-2} \del_k f \del_j f \sum_{l<m} \hat{X}^k(z_l)\hat{X}^j(z_m)
\ee
$$ +\  {1\ov 2} f^{2n-1} (\del_i \del_j f) \sum_{m=1}^{2n} \hat{X}^i(z_m)
\hat{X}^j(z_m) + \dots \big] >$$ The first term in the square
brackets is easy to evaluate  since it is the same integral as in
\ci{Lar02}. Its   contribution is $\Z_0 (T)-1$. The second term
does not contribute. The last one leads to a logarithmically
divergent term which after a renormalization gives an ambiguous
contribution \be \la{Wamb} W^{00}_{amb}= s_1 \sum_{n=1}^{\infty}
(-1)^n (e^{x^0 \ov \sqrt{2}})^{2n} {f^{2n-1} \ov 2^n} n \del^2 f
= {1\ov 2} s_1 \del^2 T {\del \Z_0  \ov \del T} \ . \ee In the
third term  we first do the integration over  the fermionic
coordinates $\th$ and use (as in the previous sections) the
symmetry of the integrand to get rid of the $\Th$ functions. This
leads to the following integral
$$ I_{2n}=-
{1 \ov (2n)!} \int \prod_{l=1}^{2n} {d \t_l \ov 2 \pi} W(1,2,\dots
2n) \ \big[\sum_{i<j} \ln |e^{i \t_i}-e^{i\t_j}|^2 + 2n\big]
 $$ \be \la{SEIn}
 \times \big[n +  \sum_{i<j} \cos (\t_i-\t_j)\big]  \ . \ee
The contribution  of the first term in the second brackets
  is proportional to \rf{In} and leads to
  \be
\la{K} N_{2n}=-  { n \ov 2^{n-1}} \big[ \sum_{m=0}^{n-2} {n-m-1 \ov
m +1}\Th(n-2) +n \Th(n-1)\big] \ .  \ee
The product of the second term in  the second brackets  in \rf{SEIn}
with the second  term in the first brackets
leads to an integral
similar to the one in
\ci{Lar02} and gives
\be \la{L} L_{2n}= - {1 \ov
2^{n-1}} n(1-n)\Th(n-1)\ .  \ee
The remaining complicated  contribution
is  evaluated in Appendix F. It reads
 \be \la{Mm} M_{2n}= {1 \ov
2^{n-1}}\bigg[ (n-1)\sum_{m=0}^{n-2} {n-m-1 \ov m+1} -{1\ov 2n}
\Theta(n-1)+ \Theta(n-2) + \delta_{n,1} \bigg] \ . \ee
Combining \rf{K},\rf{L},\rf{Mm} gives
$I_{2n}= N_{2n} + L_{2n} + M_{2n}$
in  \rf{SEIn}. Plugging it into
\rf{W001} we get
\be \la{W002} W^{00}= (\del_i
f)^2\sum_{n=1}^{\infty} (-1)^n (e^{x^0 \ov \sqrt{2}})^{2n} ({-f^2
\ov 2})^{n-1} \big[
 \sum_{m=0}^{n-2}{n-m-1 \ov m+1}\Theta(n-2) +{1\ov 2n} + (n-1)
\Th(n-2)\big]\ .  \ee
Adding
non-derivative part of  \rf{W001} we
get finally
\be \la{W003}
 W^{00}=\Z_0  -1 + (\del_i T)^2
\big[  ({1 \ov T^2 } - {T^2  \ov 2(1+\ha T^2 )^2})\ln(1+\ha T^2 )
- { T^2  \ov 2(1+\ha T^2 )^2}] \ .   \ee The 00 component of the
SET thus  becomes \be \la{T001} T^{00} =
2K \bigg[ -1 +  [ {1 \ov T^2 }\ln(1+\ha T^2 ) -
 {1 \ov 1+\ha T^2 }] (\del_i T)^2 + O((\del T)^3) \bigg] \ . \ee
 The value of the
normalization constant $K$ can be determined
by considering the limit of
$f=$const: then  $K$
can be interpreted as the tension of the original  D-brane
 $K= {1\ov 2} {\cal T}_{p}$ \ci{Lar02}.

The conservation  of SET requires
 \be \del_0 T^{00} + \del_{i} T^{i0}=0\ .  \ee
Ignoring higher-derivative terms, that leads in the present case
to the equation:\be \la{eqlin}
 {T  \ov (1+\ha T^2 )}\del^2_i  T = 0 + O((\del T)^3) \ .  \ee
 For $T= f(x_i) e^{\frac{ x^0}{\sqrt 2}}$ this
  equation is equivalent to the
 leading-order marginality condition $\del_i^2 f=0$.
 This is the expected conclusion since the conservation of the SET
 should be automatic for conformally-invariant backgrounds.
 The condition  $\del_i^2 f=0$ is solved by
  the linear background
  \be \la{line}
  f(x)= f_0 + q_i x^i \ ,  \ee
  or $T$ in  \rf{linn}.

\bigskip

Let us comment on  the physical interpretation  of the rolling
tachyon
 background with a linear spatial  profile \rf{line},\rf{linn}.
 By a global rotation we can always set
 $q_i x^i= a x, \ x=x_1$, i.e. it is sufficient to consider a
 ``one-dimensional''
inhomogeneity, $f=  q x$ (we can absorb $f_0$ into $x$).
In ref.\ci{harv,mor} it was shown that the
 spatial (time-independent)
 linear tachyon  perturbation $T(x)=qx$
results in an RG
flow for the coefficient
 $q$ from zero  to
infinity, which effectively changes
the boundary condition in the $x$-direction from the
 Neumann to the Dirichlet one.
It is natural  to expect that the target-space
time $x^0$ evolution from $-\infty$ to
$+\infty$ in the present case of
\be T=\  q(x^0) \ x\ , \ \ \ \ \ \ \ \  \ \ \ \ q(x^0) = q
e^{\frac{x^0}{\sqrt2}}   \ee
  simulates this situation in the ``on-shell'' case where
  the world-sheet theory remains
conformal throughout the time evolution.
In ref.\ci{Sen03} the
change of sign of the stress-energy
tensor for a critical value of the
tachyon  field $T_{cr}$ was associated with
 an  emergence of a codimension
one  D-brane from the rolling tachyon decay of a non-BPS D-brane.
A similar sign change happens
in  our case where the energy density $T^{00}$  in \rf{T001}
passes through zero at some value of time
(for fixed $q$ and $x$).
A particular  location of the space-time point where the sign
change occurs is  an artifact of the derivative expansion. One may
expect that this location will move to $T=  \infty$ region once
higher-derivative corrections to the stress-energy tensor  are
included.


\section{Concluding remarks
  }

The superstring partition function \rf{SF}
found in Section 4 may be interpreted as
 giving  the value
of the corresponding effective Lagrangian evaluated on the
inhomogeneous rolling tachyon background
$T= f(x_i) e^{\frac{x^0}{\sqrt 2}}$ to the second order in spatial
derivatives of the tachyon
\be \la{SFf} L= -\Z=
-\frac{1}{1 + \ha T^2 } \blp 1 + \frac{1}{1 + \ha T^2 }\bigg[ 1-
\ln(1 + \ha T^2)   + \ha  s_1 \frac{1 - \frac{3}{2}T^2 }{ 1+ \ha T^2 }
\bigg] \ (\del_i T)^2 + \dots \brp \ ,    \ee
where $s_1$ can be changed by a field redefinition.
As already mentioned in the Introduction,
this may be related to the
TDBI action \rf{nia} evaluated on the same tachyon profile
\rf{knn} by a complicated field redefinition
involving time derivatives of the tachyon.
This issue requires further study.


Expanding \rf{SFf} at
small $T$ we get
\be \la{qua}
L\approx  -1 + \ha T^2 -
\four  T^4 - (1 + \ha  s_1) (\del_i T)^2  + \dots \ .   \ee For large
$T$
 \be \la{SFF}
 L\approx - \frac{2}{ T^2 } \left[ 1
+ \frac{2}{ T^2 }( c_1  -2\ln  T)(\del_i T)^2 + \dots \right] \ , \ \ \ \ \ \ \ \
c_1 = 1 + \ln 2  -{ \textstyle \frac{3}{2}}  s_1  \ .
\ee
The direct  variation of  \rf{SFf}
over spatial tachyon profile
  does not lead to the expected leading-order
marginality (i.e. beta-function, cf.\rf{derr}) equation,
$\del^2_i f=0$; in particular, \rf{SFf}  does not have
  the  linear background \rf{line} as its exact  solution.
At the same time, the condition of conservation of
 stress-energy tensor
\rf{eqlin}
did  lead us to the correct on-shell condition
in the two-derivative approximation.

 This may look puzzling, but  has a simple explanation.
 To  be able to derive the correct equations of motion
one  needs to compute first the partition function
for the general profile function $f(x_0,x_i)$ depending also
on the  time direction. The reason is that since our tachyon
background \rf{ggg} depends on time, the time derivatives
which are  acting on
$f(x_0,x_i)$ in the full action may become acting on $T$-dependent
factors in the equation of motion, producing new terms
compared to the case where one starts with the action depending only on
$f(x_i)$.  For example, if  the action contains the term
$  T^n \del^2_0 f(x_0,x_i)$  where $T= f(x_0,x_i) e^{\frac{x^0}{\sqrt
2}}$, then in the equation of  motion we may get a term
$ \del^2_0 T^n $ which may give a non-zero contribution even after we
replace $f(x_0,x_i)$ by  $f(x_i)$.

A somewhat related comment applies to the
 expression for the stress-energy
 tensor found in Section 4.  One may wonder if  $T^{mn}$
  (or at least  its spatial components)  can
  be obtained from a  covariantization
 of  \rf{SFf}  and a
 variation over the background metric (as, e.g.,
 in \ci{Sen03,Smed03}).  This is not the case:
 considering a generalization of
  \rf{SFf}  to a constant spatial metric and taking
 derivative over it we get an expression similar but not exactly
 equal to     \rf{Tii2}.
 One possible reason for this discrepancy
 is that  tachyon action on a curved
 background  may contain terms depending on derivatives of the metric,
 e.g.,   $g(T) R+ h(T)D^i\del_i T$. Then taking
 a variation of the
action over the metric  and then setting the metric to be flat
one
may produce additional tachyon derivative terms in $T^{ij}$.\foot{
It may be of interest to
 verify this  expectation by computing the three-point
scattering amplitude for  a graviton and  two tachyons
to extract the coupling
of gravity to order $T^2$ and  thus to compare it with \rf{SFf}.}

As we discussed in the Introduction,
 there is no direct   Lorentz-covariant extension
of $Z$ computed in {\it derivative expansion}
 at the vicinity of \rf{ggg}.
Still, it may be of interest
 to  study a ``model'' covariant  action obtained from
the partition function \rf{SFf} by  simply  replacing
$\del_i T \to \del_m T$.
 Then    \rf{qua}   with $s_1=0$
will agree with  the standard quadratic tachyonic action \rf{pe}
(up to an  overall normalization
 $  {1 \ov 2}$), and so  one would reproduce the correct tachyon
  equation
 of motion to linear order in $T$.
\foot{At the same time, if one uses the replacement
 $ (\del_i T)^2 \   \to\   (\del_m T)^2 + \ha T^2  $
which  is the  identity   for the  near-mass-shell tachyon background
\rf{ggg} we have considered, then to match \rf{pe}
 one would need to set  $s_1=-1$.
In this case
the potential term and the coefficient of the $(\del_m T)^2$
term  may  get contributions also from higher-order terms  in spatial
 derivative expansion which we ignored in \rf{SF}.}
We then get a direct covariantization of
\rf{dee}
 \be\la{led} L= - V_0(T)
-  V_1(T) (\del_m T)^2  \ ,  \ee
where for $s_1=0$ in \rf{SFf}
 \be\la{vvv} V_0
= \frac{1}{1 + \ha T^2 } \ , \ \ \ \ \ \ \ \ \ \ V_1(T) = \frac{1-
\ln(1 + \ha T^2)}{(1 + \ha T^2)^2} \ . \ee
As was already discussed in the Introduction,
the change of   sign of the kinetic   function $V_1$  in \rf{led} at
$T=\pm |T_*|\approx 1.85$
suggests  development of  a  spatial inhomogeneity.
We arrived at a qualitatively similar conclusion in the
analysis of the stress-energy tensor in the previous section.
In both cases higher derivative corrections  are expected to push
the location of the transition point to $T\sim \infty$, i.e.
$x^0\to \infty$ region.

Redefining $T$ to get the standard kinetic term in the region $ 0 < |T| < |T_*|$,
i.e.  $ L= - (\del_m T')^2 - V(T')$,
one finds the potential $V(T')$  that changes from a maximal value 1  at the
 tachyonic  vacuum
at $T=T'=0$ to its minimal  value   $0.368$  at $T'(T_*)=1.04$.
Alternatively, in the region    $ |T_*| \ll  |T| < \infty $ one
may  redefine  $T$ to try connect the resulting action to the one in
\rf{deri}.  Indeed, in the large $T$ limit  the  Lagrangian \rf{SFf} or \rf{SFF}
becomes similar (but not equivalent) to
 \rf{deri} (after  a field
redefinition $T^2 \to e^{{1\ov 4} T^2}$).


Finally, let us mention again  that it would be interesting  to
 compute the exact  expression for the
superstring partition function  on some special
spatially-inhomogeneous  backgrounds  which are exactly marginal.
In particular, for   the linear profile one
  $T=(f_0 + q_i x^i) e^{x^0 \ov \sqrt 2}$  discussed above
  it is easy
to show  that the corresponding partition function contains no
2-d UV divergences to all orders in expansion in $q_i$,
suggesting that this is indeed an exactly marginal background.
 \foot{The
logarithmic singularities from correlators of $x_i$ at coinciding points
are always suppressed by  zeroes from correlators of $e^{x^0 \ov \sqrt 2}$;
in particular, the OPE of two of such operators is ``soft''.}
Computing the exact
dependence of $\Z$ on $q_i$ is equivalent to fixing its dependence
on all powers
of the gradient  $\del_i f$ in $T=f(x_i) e^{x^0 \ov \sqrt 2}$.
 Unfortunately, in contrast to the case of the simplest
``off-shell'' tachyon background $T= q_i x^i$ in \ci{mor} which
leads to a
gaussian  world-sheet theory,
it is not clear at the moment
how to compute explicitly the coefficient of  generic $q^n$ term in
expansion of  $\Z$ in powers of $q_i$.

\subsection*{Acknowledgments}

We are grateful to F. Larsen, A. Linde, H. Liu,
 R. Kallosh,    A. Naqvi
and F. Quevedo  for
informative and stimulating  discussions.
We would like also to  thank D. Kutasov and V. Niarchos for
 clarifying
discussions  and  comments
on the first version  of this paper.
 This work was supported  in part by the  PPARC SPG grant
00613. The
  work of  A.T.~was supported also  by the DOE grant
DE-FG02-91ER40690, INTAS  grant  99-1590  and the RS
Wolfson award. Part of this work was done while A.T.  was a
participant of the Superstring Cosmology Workshop at Kavli
Institute for Theoretical Physics at Santa Barbara supported in
part by the  NSF grant  No. PHY99-07949.


\renewcommand{\thesection}{A}

\setcounter{equation}{0}

\renewcommand{\theequation}{A.\arabic{equation}}

\appendix
\section{Method of orthogonal polynomials
for computing integrals in   the bosonic
case} \la{Abos}

In this Appendix we discuss some  details of the computation of
the integral which appears in eq. (19) of \cite{Lar02}:
\be\la{Lar}\int \prod_{i=1}^n \frac{d\t_i}{2 \pi}\ |\Delta(\t)|^2\
=n! \ , \ee where the Vandermonde determinant $\Delta(\t)$ is: \be
\la{Van}\Delta(\t)= \prod_{i<j}^n (e^{i\t_i}-e^{i\t_j})=
\sum_{\Pi\{i_k\}} \prod_{l=0}^{n-1} (-1)^{\Pi} e^{i \Pi(l) \t_l}
=\sum_{\Pi\{i_k\}} \prod_{k=1}^{n} (-1)^{\Pi} P^{k-1}(\l_{i_k})\ .
\ee $\Pi$  stands for all permutations  of the $\{i_k\}$
indices and the
 polynomials $P^m(\l_k) \equiv P^m(k)$
are defined  by \be   P^m(k)\equiv P^m(\l_k) = \l_k^m\ , \ \ \ \ \ \ \ \ \ \ \l_k
= e^{i \t_k}  \ , \ee with the orthogonality property \be
\int^{2\pi}_0 \frac{d\t}{2\pi} P^m (\l)  \bar{P}^l(\l) =
\de_{ml}\ . \ee
Note that in the integral of the  product
$\Delta \times\bar{\Delta}$
all cross-terms vanish due to the orthogonality of the
polynomials, so we are left with $n!$ combinations.
In other words, the only surviving permutations
are the ones which have for each
$P_m(\l_i)$ also  its complex conjugate.

Now  let us  consider the integral
$$ J_{mn}=  \int
\prod_{i=1}^n \frac{d\t_i}{2 \pi}\  |\Delta(\t)|^2 \cos
m(\t_1-\t_2)=\int \prod_{i=1}^n \frac{d\t_i}{2 \pi} \ \l_1^m
\Delta(\t) \bar{\l}_2^m \bar{\Delta}(\t) \ ,  $$
and prove the following general result
 \be \la{VanCos}
J_{mn} (m>n-1) =0 \ , \ \ \ \ \ \ \ \ \
J_{mn} (m\leq n-1) =-(n-m)(n-2)! \ ,
\ee
where we used in  the symmetry of $1 \to 2$.
First, let us  note the following relations
$$\l_1^m \{ \dots P^l(\l_1) \dots P^{l'} (\l_2) \dots \} (-1)^{\Pi}=
\{ \dots P^{l+m}(\l_1) \dots P^{l'} (\l_2) \dots \} (-1)^{\Pi}$$
 \be \bar{\l}_2^m \{ \dots \bar{P}^k(\bar{\l}_1)
\dots \bar{P}^{k'} (\bar{\l}_2) \dots \} (-1)^{\Pi'}=
 \{ \dots \bar{P}^{k}(\bar{\l}_1) \dots \bar{P}^{k'+m}
(\bar{\l}_2) \dots \} (-1)^{\Pi'} \ee
 Due to the orthogonality of
the polynomials we must have $l+m=k$ and $l'=k'+m$ with all other
$n-2$  factors  $\l_{i_k}$  distributed the same way in
$\Delta(\t)$ and $\bar{\Delta}(\t)$. This implies that $\Pi'$ is
an odd permutation of $\Pi$ and there are $(n-2)!$  ways to
distribute  $\l_{i_k}, \ i \neq l,l'$. Finally, the above
relations imply that $m+1 \leq l'=l+m \leq n$ and this gives $n-m$
possible values of $l$. Also,  if $m>n-1$ we get vanishing result
since $P^{m+l}(\l_{i_l})$ cannot be paired with its complex
conjugate in $\bar{\Delta}(\t)$. This leads to (\ref{VanCos}).

One can easily check that  (\ref{VanCos})  can be used to reproduce (\ref{tak})
and also eq.
(28) in \cite{Lar02}, i.e.
\be
\int
\prod_{i=1}^n \frac{d\t_i}{2 \pi}\  |\Delta(\t)|^2
[ n + 2\sum^n_{i < j} \cos (\t_i-\t_j)] = n! \ .
\ee

\renewcommand{\thesection}{B}

\setcounter{equation}{0}

\renewcommand{\theequation}{B.\arabic{equation}}
\section { Path ordered integral of an antisymmetric function}\label{AP}

 In this appendix we will show how   the path ordered integral $\int [d\t]_{2n }$
(see \rf{mea}) of
an antisymmetric integrand $A(1,...,2n)$  can be converted into  an  integral
 without
path-ordering.
As  is well known, for a symmetric integrand one has
 \be \la{opp}
 \int [d\t]_{2n} \ S(1,...,2n)= \frac{1}{(2n)!} \int    \prod_{i=1}^{2n} \frac{d \t_i}{2\pi}
   \  S(1,...,2n)\  .
\ee
 We can
use this identity to show that for a fully antisymmetric integrand
one gets
\be \int [d\tau]_{2n}\   A(1,\dots,2n)= \frac{1}{(2n)!} \int
\prod_{i=1}^{2n}  \frac{d\t_i}{2 \pi}\ \prod_{i<j}^{2n}   \ep(i,j) \ A(1,
\dots,2n)\ . \ee
 To prove \rf{opp} in  the symmetric function  case we
use that $1=\Th(i,j)+ \Th(j,i)$ to  write   \be
\frac{1}{(2n)!} \int \prod_{i=1}^{2n} \frac{d\t_i}{2 \pi}\ S(1,...,2n)=
\frac{1}{(2n)!} \int \prod_{i=1}^{2n} \frac{d\t_i}{2 \pi}\prod^{2n}_{i<j}\
[\Th(i,j)+ \Th(j,i)] \ S(1,\dots,2n) \ . \ee
 Expanding the product
of Theta functions on the r.h.s. we get $2^{\frac{2n(2n-1)}{2}}$
terms out of which only $(2n)!$  have non-circular orderings, i.e.
give non-vanishing contributions.
Each non-vanishing  term has a string $n(2n-1)$ Theta-function
factors but only $2n-1$  are needed to  get path ordering in the
integral over $n$ points.
 The remaining  act as constrains which are automatically
satisfied for each of the $(2n)!$  non-vanishing orderings.


Now to prove  (\ref{iden}) we start from its r.h.s.
and  that  $\ep(i,j)= \Th(i,j)- \Th(j,i)$.
Then    going   through the same arguments as above  we will
find all path orderings but with plus or minus sign depending on
whether they are odd or even permutations of $\{1,2,\dots, 2n\}$.
Using the antisymmetry of $A(1,...,2n)$  we can show that they all are equal to
the same path ordered  integral  $ \int \prod_{i=1}^{2n} \frac{d\t_i}{2 \pi}  \Th(1,2)\dots \Th(2n-1,2n)$,
and that constitutes the proof of  (\ref{iden}).

\renewcommand{\thesection}{C}

\setcounter{equation}{0}

\renewcommand{\theequation}{C.\arabic{equation}}

\section{Properties of W(1,\dots,2n)  }\label{PZ}

Here we shall comment on some properties of the function $W$
defined by \rf{www}. i.e. \be \la{W} W(1,\dots, 2n)= \sum_{\P}
(-1)^{\P} \frac{\prod^{2n}_{i<j} D(i,j)}{D(i_1,i_2)\dots
D(i_{2n-1},i_{2n})}\ . \ee The summation is over only the
$(2n-1)!!$  permutations of pairs of indices with  $(-1)^{\P}$ as
the symmetry factor of permuting indices between pairs but always
in an ordered manner from lower to higher,  i.e.
$\{(1,2),\dots,(2n-1,2n)\} \to \{(1,2n),\dots, (2,2n-1)\}$, etc.
If we sum also  over exchanges in pairs $(i_1,i_2)\to (i_2,i_1)$
and use the fact that $D(i_1,i_2)=- D(i_2,i_1)$ then  we get $2^n$
extra terms and we need to divide by this factor: \be W(1,\dots,
2n)= \frac{1}{2^n}\sum_{\PP} (-1)^{\PP} \frac{\prod^{2n} _{i<j}
D(i,j)}{D(i_1,i_2)\dots D(i_{2n-1},i_{2n})}\ .  \ee Here $\PP$
stands for permutations including also
 interchanges in each pair. There are then $(2n-1)!!\
2^n$ terms.
Next,  we can sum over the  $n!$  interchanges of pairs among
themselves. This way we can write $W(1,\dots,2n)$ as a sum over
all $(2n)!$ permutations.
The final form we get is:
\be\la{Z}
W(1,\dots, 2n)= \frac{1}{2^nn!}\sum_{\Pi} (-1)^{\Pi}
\frac{\prod^{2n}_{i<j} D(i,j)}{D(i_1,i_2)\dots D(i_{2n-1},i_{2n})}
\ ,
\ee
where $\Pi$ stands for all permutations.

\renewcommand{\thesection}{D}

\setcounter{equation}{0}

\renewcommand{\theequation}{D.\arabic{equation}}

\section{Details of integral evaluation in Section 3.1}\label{AZ}


To compute the superstring partition function
 to zero order in expansion in derivatives (\ref{SPF1}) one
needs to find  the constant
(non-oscillating) part of $W(1,\dots,2n)$. We can rewrite the
integrand in   (\ref{SPF1}) as:
$$ W(1,\dots,2n)= \sum_{\P} (-1)^{\P}
(\bar{P}^n(i_1) \bar{P}^{n-1}(i_2)\dots \bar{P}^n(i_{2n-1})
\bar{P}^{n-1}(i_{2n}))$$ \be\times
\frac{\Delta(\t)}{(1-\bar{P}(i_1)P(i_2)) \dots
(1-\bar{P}(i_{2n-1})P(i_{2n}))} \ . \ee
 Expanding the
denominators in power series we get
 $$  W(1,2,\dots,2n)=\sum_{\P} (-1)^{\P}
(\bar{P}^n(i_1) \bar{P}^{n-1}(i_2)\dots \bar{P}^n(i_{2n-1})
\bar{P}^{n-1}(i_{2n}))$$ \be \la{Z01}\times
[\Delta(\t)\prod_{k=1}^n \sum_{m_k} (
\bar{P}^{m_k}(i_{2k-1})P^{m_k}(i_{2k}))]\ . \ee From the second
term (in square brackets) we need to pick up the complex conjugate
to the first one.

Let us first consider  the $\{1,2,\dots,2n\}$ ordering in the sum
over permutations. The basic observations   which we use in  our
computation are: (a) in $\Delta(\t)$ each ordering of powers of
the polynomials appears only once and each polynomial $P(i_k)$ has
a different power compared to others; (b) if  from
$\prod_{k=1}^n\sum_{m_k} \dots$ we pick up  the term
$\bar{P}(i_1)^s P(i_2)^s$ ($m_1=s$)  then from $\Delta(\t)$ we
must pick up  the term $P(i_1)^{n+s} P(i_2)^{n-s-1}$ with $n+s
\leq 2n-1, \ 0\leq n-1-s$; (c) since each power $0,\dots,  2n-1$
appears in $\Delta(\t)$ only once,
  all  possible  powers $m_k= 0,
\dots,  n-1$ are different from each other; (d) there are  $n!$
permutations
 distributing  the distinct $m_k$ powers over the pairs of
polynomials that appear in (\ref{Z01})  and we need only to
determine the relative signs of them.

Let $S^0=P^{2n-1}(1) P^{2n-2}(2) \dots P^0(2n)$ be a reference
configuration in $\Delta(\t)$ with $(-1)^{\Pi(S^0)}$ =+1. If we
want construct a configuration with a given distribution of powers
$\{m_k\}$ then we proceed from the $S^0$ as follows: (a) Permute
the pair of indices $(i_{2k-1}, i_{2k})$ which has the maximum
$m_k=n-1$ to the position such that $P(i_{2k-1})$ has power
$n+m_k=2n-1$ and $P(i_{2k})$ has power $2n-2$.  For this step we
need an even number of permutation since we are permuting a pair.
(b) Now take the $i_{2k}$ index $2m_k=2n-2$ positions to the right
to power zero. ( c) Follow the same procedure for all other pairs
of indices in descending order of powers. In this way we can
construct all strings of polynomials $P(i_1)^{2n-1} P(i_2)^{2n-2}
\dots P(i_{2n})^0$ by making an even number of permutations. We
conclude that all the terms in  $\Delta(\t)$ which are picked up by
this procedure have $(-1)^{\Pi(S^{\{m_k\}})}=(-1)^{\Pi(S^0)}=+1$.
Therefore,  they all add up to give a contribution  of  the term
with $(1,2,\dots, 2n)$ ordering in $W(1,\dots, 2n)$ to be
  $n!$.

The final step is to find  the contribution of the remaining
$\P\{1,\dots,2n\}$ terms in $W$. This is easily done if we notice
that we can write $\Delta(\t)= \prod_{i<j} (P(i)-P(j))$ =
$(-1)^{\P\{1,\dots,2n\}}$ $ \prod_{\P(i<j)} (P(i)-P(j))$ . Here
$\P(i<j)$ means the  permutation of the initial ordering,  i.e. if
$\P (1234...)= (1324...)$ then
$\P(i<j)=\P(1<2<3<4...)=(1<3<2<4...).$ We then proceed  in the
same way
 with the proof as above to  find that the contribution is
equal to $(-1)^{\P} n!$. The final result is then: \be \la{Z02}
\int \prod_{i=1}^{2n} \frac{d \t_i}{2\pi}W(1,\dots,2n)=
\sum_{\P}(-1)^{\P} [(-1)^{\P}n!]= (2n-1)!! \ n! \ . \ee

\renewcommand{\thesection}{E}

\setcounter{equation}{0}

\renewcommand{\theequation}{E.\arabic{equation}}

\section{Details of integral evaluation in Section 3.2}\label{EE}

In this Appendix we will discuss computation
of  the integral $C_n$ from
(\ref{Z1C}). First,  we write the cosine as $P(2)^M \bar{P}(1)^M$
where we used the symmetry of the integrand under $1\to 2$. In the
definition of $W(1,\dots,2n)$ there are two kinds of terms: (i)
those where $D(1,2)$ appears in the denominator of the fractions
in \rf{SPF1} and (ii) those where $D(1,*)D(2,*)$ appears instead,
where $*$ is any other index from $\{ 3,4 ,\dots \}$.

The first case is easily worked out. Expanding the denominators we
get: $$ (-1)^{\P(12 \dots )}\bar{P}(1)^M P(2)^M (\bar{P}^n(i_1)
\bar{P}^{n-1}(i_2)\dots \bar{P}^n(i_{2n-1})
\bar{P}^{n-1}(i_{2n}))$$
 \be \times\ \Delta(\t)\prod_{k=1}^n
\sum_{m_k} ( \bar{P}^{m_k}(i_{2k-1})P^{m_k}(i_{2k})) \ . \ee As in
the computation of $\Z_0$   the integral will project from
$\Delta(\t)$ those strings of polynomials which are conjugates to
the rest of the integrand for each permutation of powers
$\{m_k\}\equiv \{0, \dots, n-1\}$. There are two constrains
imposed by the minimal (0) and maximal $(2n-1)$ power in
$\Delta(\t)$: $1 \leq M \leq N-1$ and $0\leq m_1 \leq n-1-M$.
Therefore,  there are exactly $(n-M)$ possible values of $m_1$ and
the remaining $n-1$ powers $m_k$ can be distributed in $(n-1)!$
ways over the remaining pairs of indices. Each of the $(n-M)
\times (n-1)!$ terms has the same sign, $(-1)^{\P\{12\dots\}}$,
since the required strings of polynomials in $\Delta(\t)$ are
constructed just like in the zeroth order calculation in Appendix
\ref{AZ}. The total contribution of  these terms is found after
summing over all $(2n-3)!!$ combinations in $W(1,\dots,2n)$ which
have $D(1,2)$ in the denominator:
$$ Q^{(1)}_n(M) = \sum_{ \{ (12), \dots \} }(n-M) \times (n-1)!\
[(-1)^{\P(1,2,\dots,2n)}]^2$$ \be \la{I1} = \   (2n-3)!!\  (n-M)\
(n-1)! \ . \ee The second type  of terms corresponds to the
remaining $(2n-2) (2n-3)!!$ ways of pairing indices 1 and 2 with
any other index except  with  each other. Here a typical term is:
$$(-1)^{\P(1,i_3,2,i_4 \dots )}\bar{P}(1)^M P(2)^M (\bar{P}^n(i_1)
\bar{P}^{n-1}(i_3) \bar{P}^{n}(i_2) \bar{P}^{n-1}(i_4) \dots
\Delta(\t) $$ \be \times \sum_{m_1} (
\bar{P}(1)^{m_1}P(3)^{m_1})\sum_{m_2} (
\bar{P}(2)^{m_1}P(4)^{m_2}) \prod_{k=3}^n \sum_{m_k} (
\bar{P}^{m_k}(i_{2k-1})P^{m_k}(i_{2k})) . \ee The determinant
imposes again the constrains: $1 \leq M \leq N-1$ and $0\leq
m_{1,2} \leq n-1-M$. The remaining $n-2$ powers $m_k$ are
distributed between  the remaining $n-2$ pairs. It is obvious from
the analysis of Appendix \ref{AZ}, that the two members of a given
pair of polynomials are split symmetrically with respect to the
pair with powers $(n,n-1)$, i.e.
$P(i_{2k-1})^{n+m_k}...P(i_{2k'-1})^{n}P(i_{2k'})^{n-1}$ $\dots
P(i_{2k})^{n-1-m_k}$. In our case this implies that we must have
$m_2=m_1+M$. This means that the pairs $(1,i_4)$ and $(2,i_3)$ are
the ones with symmetric polynomial powers above and below the pair
of polynomials with powers $(n,n-1)$ respectively. We would like
to follow the method of Appendix \ref{AZ} to construct the
required strings of polynomials from the reference one, $S^0$. As
a first step we have to permute $(1,i_3,2,i_4,...) \to
(1,i_4,2,i_3...)$ in $\Delta(\t)$ resulting in a minus sign. The
rest goes as in the first case with all terms in each
$\P\{1,i_3,2,i_4 \dots \}$ contributing the same quantity
$-(-1)^{\P\{1,i_3,2,i_4 \dots \}}(n-M)(n-2)!$ and finally: \be
\la{I2} Q^{(2)}_n(M) =- (2n-2) (2n-3)!! (n-M) (n-2)! \ee Adding
the two expression in (\ref{I1}) and (\ref{I2}) for each $M$ leads
to the result in (\ref{In}).

Let us consider  an explicit example which will make the procedure more
transparent. Take the case $n=3$ and let us look at the following
term of the type  (ii): $$(-1)^{\P(1,3,2,4,5,6)}\bar{P}(1)^M P(2)^M
(\bar{P}^3(1) \bar{P}^{2}(3) \bar{P}^{3}(2)
\bar{P}^{2}(4)\bar{P}^{3}(5) \bar{P}^{2}(6) $$ \be \la{example}
\times  \Delta(\t) ( \bar{P}(1)^{m_1}P(3)^{m_1})(
\bar{P}(2)^{m_1}P(4)^{m_2})  ( \bar{P}^{m_3}(5)P^{m_3}(6))  \ ,  \ee
where we have suppressed the $m_1,m_2,m_3,M$ summations.
Here  the powers $x_i$  of each polynomial  $P^{x_i}(i)$ are: $$
({x_1},{x_3},{x_2},{x_4},{x_5},{x_6})=
 (3+M+m_1,2-m_1,3+m_2-M,2-m_2,3+m_3,2-m_3)\ , $$
 where we used the
identity $P(i)^m=\bar{P}(i)^{-m}$. Since in $\Delta(\t)$ there are
only polynomials with degrees 0 to 5, this implies that $1\leq
M\leq 2$ and $0\leq m_{1,2} \leq 2-M$. Moreover,  $0\leq m_3 \leq
2$. We see that the last pair $(5,6)$ has powers symmetrically
$m_3$ steps higher and $m_3$ steps lower with respect to the pair
which has powers $(3,2)$. Also,  from the powers of $(1,3,2,4)$ we
conclude that two of them (those of $P(1),P(2)$) are 3 or higher
and the other two (of $P(3),P(4)$) are 2 or lower.

Now we have two possibilities: (i)  $P(1)$ and $P(3)$ have powers
symmetric with respect to the center (and the same for $P(2)$ and
$P(4)$) implying  $3+M+m_1=3+m_1$ which is impossible since $M\geq
1$,  and (ii)  $P(1)$ and $P(2)$ have powers symmetric with
respect to the center (and the same for $P(2)$ and $P(3)$) which
implies $3+M+m_1=3+m_2$.  Then  our string of polynomials
becomes:$$\bar{P}(1)^{3+M+m_1}\bar{P}(3)^{2-m_1}\bar{P}(2)^{3+m_1}
\bar{P}(4)^{2-M-m_1}\bar{P}(5)^{3+m_3}\bar{P}(6)^{2-m_3} $$
 Now there
are two possibilities: $M=1$ and one for $M=2$ which lead to
distinct powers $\bar x_i= - x_i$ of
 each of the six polynomials ${\bar P}^{\bar x_i} $ (i)
 \be
\begin{array}{|c||c|c|c|c|}
\hline M & m_1 & m_3 &
({\bar x_1},{\bar x_4},{\bar x_2},{\bar x_3},{\bar x_5},{\bar x_6})
&
(-1)^{\Pi(S^{\{m_k\}})}
\\
\hline 1&0&2&(4,1,3,2,5,0)&+1
\\
\hline 1&1&0&(5,0,4,1,3,2)&+1
\\
\hline 2&0&1&(5,0,3,2,4,1)&+1
\\
\hline
\end{array}\nonumber
\ee
In the last column  we have given
the sign of each string of
polynomials in $\bar{\Delta'}(\t)$ =$ (-1)^{\P\{1,4,2,3,5,6\}}$ $
\prod_{\P(i<j)} (\bar{P}(i)-\bar{P}(j)) $ with reference to the
$S'=\bar{P}(1)^{5}\bar{P}(4)^{4}\bar{P}(2)^{3}
\bar{P}(3)^{2}\bar{P}(5)^{1}\bar{P}(6)^{0} $ string. The signs of
the strings of polynomials of the fourth column in the table are
all positive relative to $S'$.  One  can show this by using the
method of Appendix \rf{AZ}. We have to permute the indices of the
reference string $S'$ an even number of times to construct any of
the strings in the fourth column of the table, e.g.,   for the third
entry in the table above:
$\bar{P}(1)^{5}\bar{P}(4)^{4}\bar{P}(2)^{3}
\bar{P}(3)^{2}\bar{P}(5)^{1}\bar{P}(6)^{0} \to $
$\bar{P}(1)^{5}\bar{P}(2)^{4}\bar{P}(3)^{3}
\bar{P}(5)^{2}\bar{P}(6)^{1}\bar{P}(4)^{0}$

\noindent
$ \to$
$ \bar{P}(1)^{5}\bar{P}(5)^{4}\bar{P}(6)^{3}
\bar{P}(2)^{2}\bar{P}(3)^{1}\bar{P}(4)^{0}$
 $\to \bar{P}(1)^{5}\bar{P}(5)^{4}\bar{P}(2)^{3}
\bar{P}(3)^{2}\bar{P}(6)^{1}\bar{P}(4)^{0}   . $ This is the desired
result,  and if we use the fact that
$\Delta(\t)=(-1)^{\P\{1,4,2,3,5,6\}}\Delta'(\t)$ in \rf{example} we
get an overall minus sign. There are 12 such cases as the one we
studied above. Therefore,  $Q^{(2)}_3(1)=-12\times 2=-24$ and
$Q^{(2)}_3(2)=-12\times 1=-12$, in agreement with  the general expression
\rf{I2}.

\renewcommand{\thesection}{F}

\setcounter{equation}{0}

\renewcommand{\theequation}{F.\arabic{equation}}

\section{Details of  computation of  stress-energy tensor} \la{SET}

The computation of the  stress-energy  tensor uses the same
technology as developed in the previous appendices. The
computation of the $M_{2n}$ integral \rf{Mm} in Section 4 is quite
lengthy and here we shall give a short account of some
intermediate  results.

The integral we  are to compute is
 \be M_{2n}= -{1 \ov (2n)!}
\int \prod_{l=1}^{2n} {d \t_l \ov 2 \pi} W(1,2,\dots 2n)\
\sum_{i<j} \ln |e^{i \t_i}-e^{i\t_j}|^2  \  \sum_{i<j} \cos
(\t_i-\t_j) \ .  \ee Due to the symmetry under interchange of
 the integrations points it can be written
as: $$ M_{2n}= -{1 \ov (2n)!} \int \prod_{l=1}^{2n} {d \t_l \ov 2
\pi} W(1,2,\dots 2n)\  \sum_{i<j} \ln |e^{i \t_1}-e^{i\t_2}|   $$
\be \la{M} \times 2\big[ \cos (\t_1-\t_2)+ 2(2n-2)\cos
(\t_1-\t_3)+ (n-1)(2n-3)  \cos (\t_3-\t_4)\big] \ . \ee In order
to compute the integral for each of the three terms in the
brackets we expand the logarithm in a series of cosines.

Then for each
 term  we will  need to distinguish terms in
$W(1,2 \dots 2n)$ \rf{W} depending on how the special points
$1,2,3,4$ of the logarithm and $\cos (\t_i-\t_j)$ appear in the
$D(ij)$ of the denominator.
In the table that follows we present
the results for each type  of terms.
The first column contains the
structure of the denominator for the terms selected. The second
column gives
the  product of orthogonal polynomials selected from the
cosine function. The third column  gives
the number of independent
configurations of a given type in $W$.
 The fourth column contains
the result of  integration computed in the same way  as the
derivative corrections in the partition function.

In the expansion of the logarithm
we  use the symmetries of the
rest of the integrand to write: $ \ln |e^{i \t_1}-e^{i\t_2}|=
\sum_{M=0}^{\infty} {\cos [(M+1)( t_1 -t_2)] \ov M+1}=
\sum_{M=0}^{\infty} {\bar{P}(1)^{M+1} P(2)^{M+1} \ov M+1}$.
Summation over the integer variable $M$ is always
implied. Also, we use the convention $D(ij)=D(i,j)$ and
  $\Theta(k)=\Theta(n-k)=
1, $ if $n>k$ or $=0$ if $n<k$.

Combining  the  results given in the tables below   we get the
final expression  \rf{Mm} for the integral in   \rf{M}.

\paragraph {\bf I: $\ln |e^{i \t_1}-e^{i\t_2}| \cos (\t_1-\t_2)$}

\be \small
\begin{array}{|c||c|c||c|}
\hline Denominator &Polynomials &Combinations & Integral
\\
\hline D(12)& \bar{P}(1) P(2) &(2n-3)!!& -{1 \ov 2} (n-1)!
(n-2-M)\Theta(n-3-M)  \Theta(3)
\\
\hline D(1*)D(2*)& \bar{P}(1) P(2) &(2n-2)(2n-3)!!& +{1 \ov 2}
(n-2)! (n-2-M)\Theta(n-3-M)   \Theta(3)
\\
\hline D(12)& \bar{P}(2) P(1) &(2n-3)!!& -{1 \ov 2} ((n-1)!
(n-M)\Theta(n-1-M)\Theta(2)-\delta_{n,1}\delta_{M,0})
\\
\hline D(1*)D(2*)& \bar{P}(2) P(1) &(2n-2)(2n-3)!!& {1 \ov 2}(
-n^2(n-2)!\Theta(2)\delta_{M,0}
\\
& & &+(n-2)! (n-M)\Theta(n-1-M) \Theta(2))
\\
\hline
\end{array}\nonumber
\ee

\paragraph {\bf II: $\ln |e^{i \t_1}-e^{i\t_2}| \cos (\t_1-\t_3)$}

\be \small
\begin{array}{|c||c|c||c|}
\hline Denominator &Polynomials &Combinations & Integral
\\
\hline D(13)D(2*)& \bar{P}(1) P(3) &(2n-3)!!& +{1 \ov 4} (n-2)!
(n-2-M)\Theta(n-3-M)  \Theta(3)
\\
\hline D(13)D(2*)& \bar{P}(3) P(1) &(2n-3)!!& +{1 \ov 4} (n-2)!
(n-1-M)\Theta(n-2-M)  \Theta(2)
\\
\hline D(13)D(2*)& \bar{P}(2) P(3) &(2n-3)!!& +{1 \ov 4} (n-2)!
(n-2-M)\Theta(n-3-M)  \Theta(3)
\\
\hline D(13)D(2*)& \bar{P}(3) P(2) &(2n-3)!!& +{1 \ov 4}( (n-2)!
(n-1-M)\Theta(n-2-M)|_{M \neq 0} \Theta(2)
\\
&&&-(n-1)^2 \delta_{M,0} \Theta(2))
\\
\hline D(23)D(1*)& \bar{P}(1) P(3) &(2n-3)!!& +{1 \ov 4} (n-2)!
(n-2-M)\Theta(n-3-M)  \Theta(3)
\\
\hline D(23)D(1*)& \bar{P}(3) P(1) &(2n-3)!!& +{1 \ov 4}( (n-2)!
(n-1-M)\Theta(n-2-M)|_{M \neq 0} \Theta(2)
\\
&&&-(n-1)^2 \delta_{M,0} \Theta(2))
\\
\hline D(23)D(1*)& \bar{P}(2) P(3) &(2n-3)!!& +{1 \ov 4} (n-2)!
(n-1-M)\Theta(n-2-M)  \Theta(2)
\\
\hline D(23)D(1*)& \bar{P}(3) P(2) &(2n-3)!!& +{1 \ov 4} (n-2)!
(n-2-M)\Theta(n-3-M)  \Theta(3)
\\
\hline D(12)D(3*)& \bar{P}(1) P(3) &(2n-3)!!& +{1 \ov 4} (n-2)!
(n-2-M)\Theta(n-3-M)  \Theta(3)
\\
\hline D(12)D(3*)& \bar{P}(3) P(1) &(2n-3)!!& +{1 \ov 4} (n-2)!
(n-1-M)\Theta(n-2-M) \Theta(2)
\\
\hline D(12)D(3*)& \bar{P}(2) P(3) &(2n-3)!!& +{1 \ov 4} (n-2)!
(n-1-M)\Theta(n-2-M)  \Theta(2)
\\
\hline D(12)D(3*)& \bar{P}(3) P(2) &(2n-3)!!& +{1 \ov 4} (n-2)!
(n-2-M)\Theta(n-3-M)  \Theta(3)
\\
\hline D(1*)D(2*)D(3*)& \bar{P}(1) P(3) &(2n-4)(2n-3)!!& -{1 \ov
4} 2(n-3)! (n-2-M)\Theta(n-3-M)  \Theta(3)
\\
\hline D(1*)D(2*)D(3*)& \bar{P}(3) P(1) &(2n-4)(2n-3)!!& -{1 \ov
4}( 2(n-2)! (n-1-M)\Theta(n-2-M)|_{M \neq 0} \Theta(3)
\\
&&&- (n-1)! \delta_{M,0}\Theta(3))
\\
\hline D(1*)D(2*)D(3*)& \bar{P}(2) P(3) &(2n-4)(2n-3)!!& -{1 \ov
4}( 2(n-2)! (n-1-M)\Theta(n-2-M)|_{M \neq 0} \Theta(3)
\\
&&&- (n-1)! \delta_{M,0}\Theta(3))
\\
\hline D(1*)D(2*)D(3*)& \bar{P}(3) P(2) &(2n-4)(2n-3)!!& -{1 \ov
4} 2(n-2)! (n-2-M)\Theta(n-3-M)  \Theta(3)
\\
\hline
\end{array}\nonumber
\ee

\paragraph {\bf III: $\ln |e^{i \t_1}-e^{i\t_2}| \cos (\t_3-\t_4)$}

\be \small
\begin{array}{|c||c|c||c|}
\hline Denominator &Polynomials &Combinations & Integral
\\
\hline D(12)D(34) & \bar{P}(3) P(4)& (2n-5)!! &  -(n-2) (n-2)!
(n-1-M) \Theta(n-2-M) \Theta(2)
\\
\hline D(13)D(24) & \bar{P}(3) P(4)& (2n-5)!! & + (n-2)! (n-2-M)
\Theta(n-3-M) \Theta(3)
\\
\hline D(14)D(23) & \bar{P}(3) P(4)& (2n-5)!! & + (n-2)!(
(n-1-M) \Theta(n-2-M) \Theta(2)
\\
&&&-(n-1)^2 \delta_{M,0} \Theta(2))
\\
\hline D(1*)D(2*)D(34) & \bar{P}(3) P(4)& (2n-4)(2n-5)!! &  +
(n-3)!((n-2) (n-1-M) \Theta(n-2-M) \Theta(3) \\
&&& (n-2-M) \Theta(n-3-M)\Theta(3))
\\
\hline D(3*)D(4*)D(12) & \bar{P}(3) P(4)& (2n-4)(2n-5)!! &  +
(n-3)!((n-2) (n-1-M) \Theta(n-2-M) \Theta(3) \\
&&& (n-2-M) \Theta(n-3-M)\Theta(3))
\\
\hline D(1*)D(3*)D(24) & \bar{P}(3) P(4)& (2n-4)(2n-5)!! &  -
2(n-3)! (n-2-M) \Theta(n-3-M)\Theta(3)
\\
\hline D(3*)D(4*)D(12) & \bar{P}(3) P(4)& (2n-4)(2n-5)!! &  -
2(n-3)! (n-2-M) \Theta(n-3-M)\Theta(3)
\\
\hline D(1*)D(4*)D(23) & \bar{P}(3) P(4)& (2n-4)(2n-5)!! &  -
(n-3)!((n-1-M) \Theta(n-2-M)  \\
&&&+(n-2-M) \Theta(n-3-M) -(n-1)^2 \delta_{M,0})\Theta(3)
\\
\hline D(3*)D(4*)D(12) & \bar{P}(3) P(4)& (2n-4)(2n-5)!! &  -
(n-3)!((n-1-M) \Theta(n-2-M)  \\
&&&+(n-2-M) \Theta(n-3-M) -(n-1)^2 \delta_{M,0})\Theta(3)
\\
\hline D(1*)D(2*)\times & \bar{P}(3) P(4)& (2n-4)(2n-6)\times & -
(n-4)!((n-3)(n-1-M) \Theta(n-2-M)  \\
D(3*)D(4*)&&(2n-5)!!&-4(n-2-M) \Theta(n-3-M) +(n-1)^2
\delta_{M,0})\Theta(4)
\\
\hline
\end{array}\nonumber
\ee


\end{document}